\documentclass[aps,prb,twocolumn,showpacs]{revtex4}
\usepackage{dcolumn}
\usepackage{amsmath}
\usepackage{amsfonts}
\usepackage{amssymb}
\usepackage{float}
\usepackage{graphicx}
\usepackage{bm}

\begin{document}

\title{Electron-phonon interaction in Graphite Intercalation Compounds}
\author{Lilia Boeri$^1$, Giovanni B. Bachelet$^{2}$, Matteo Giantomassi$^{3}$ and Ole K. Andersen$^1$ }
\affiliation{$^1$Max-Planck Institut f\"{u}r Festk\"{o}rperforschung, Heisenbergstr. 1,
D-70569 Stuttgart, Germany\\
$^2$INFM Center for Statistical Mechanics and Complexity and Dipartimento di
Fisica, Universit\`a di Roma ``La Sapienza'', Piazzale A. Moro 2, 00185
Roma, Italy\\
$^3$Unit\'e de Physico-Chimie et de Physique des Mat\'eriaux, Universit\'e
Catholique de Louvain, place Croix du Sud 1, B-1348 Louvain-la-Neuve, Belgium%
}
\date{\today }

\begin{abstract}
Motivated by the recent discovery of superconductivity in Ca- and
Yb-intercalated graphite (CaC$_{6}$ and YbC$_{6}$) and from the ongoing
debate on the nature and role of the interlayer state in this class of
compounds, in this work we critically study the electron-phonon properties
of a simple model based on primitive graphite. We show that this model
captures an essential feature of the electron-phonon properties of the
Graphite Intercalation Compounds (GICs), namely, the existence of a strong
dormant electron-phonon interaction between interlayer and $\pi ^{\ast }$
electrons, for which we provide a simple geometrical explanation in terms of
NMTO Wannier-like functions. Our findings correct the oversimplified view
that nearly-free-electron states cannot interact with the surrounding lattice,
and explain the empirical correlation between the filling of the interlayer band 
and the occurrence of superconductivity in Graphite-Intercalation Compounds.
\end{abstract}

\maketitle

\section*{Introduction}

Superconductivity in Ca- and Yb-intercalated graphite (CaC$_{6}$ and YbC$%
_{6} $), with $T_{c}$ of respectively 11.5 and 6.5$\,$K, was discovered~\cite%
{YbC6:weller:syn,CaC6:emery:syn} five years later than in the
pseudo-graphite MgB$_{2},$ where $T_{c}$=40$\,$K.~\cite{MgB2_expt1}
Superconductivity in MgB$_{2}$ is understood to be caused by the strong
interaction between a few zone-centered holes in the B$\,sp^{2}$ $\sigma $%
-bonding band and a few zone-centered bond-stretching phonons, which
themselves get softened by the interaction;\cite%
{MgB2_theory,Pickett,Kong,BoeriDiamond} 
the $\pi $ bands are also at the Fermi level, 
but they  couple to phonons only weakly.
What makes MgB$_{2}$ special is
that the electrostatic and covalent attraction of the B $p_{z}\,\pi $
electrons by the intercalated Mg$^{++}$ ions is so strong that the center
of the $\pi $ bands is lowered \emph{below} the top of the $\sigma $ band,
whereby \emph{hole}-doping takes place into this bonding band.  An equally simple and elegant understanding of the origin of superconductivity in the strongly
\emph{electron}-doped CaC$_{6}$ and YbC$_{6}$ is still missing. Csanyi \emph{%
et al.}~\cite{GIC:csanyi:band} have empirically related the appearance of
superconductivity in Graphite-Intercalation Compounds (GICs) to the filling
of a free-electron-like interlayer (IL) band,~\cite%
{GIC:posternak:band,GIC:fauster:band} which is present, albeit empty, in
pure graphite; and proposed that plasmon-mediated superconductivity could
take place in these compounds.
 Recent experiments~%
\cite{CaC6:jskim:Cp,CaC6:lamura:penetration} and \emph{ab-initio}
calculations~\cite%
{GIC:mazin:band,CaC6:calandra:band,CaC6:jskim:press,RMTACaC6_pressure}
agree, instead, that the conventional e-ph coupling suffices to explain the
superconductivity in CaC$_{6}$ and YbC$_{6}$, although some puzzling issues
remain.\cite{GICs:Mazin:review}

It seems therefore  that in graphite, 
if the Fermi surface has only $\pi $ sheets,  the e-ph coupling is weak 
and $T_{c}\lesssim 1\,$K, whereas, if also $\sigma $ or IL sheets are present,
 the e-ph coupling increases enough as to yield much higher $T_{c}$'s. 
Mazin \emph{et al.}~\cite{GIC:mazin:band} 
and Calandra \emph{et al.}~\cite{CaC6:calandra:band} 
argued that the main role in the e-ph coupling is played by electronic and
vibrational 
states associated with the presence of the intercalant ion, and, generally speaking, this is true: without the
intercalant, graphite is not a superconductor. However, a close inspection
of the e-ph spectral functions $\alpha ^{2}F(\omega )$ of CaC$_{6}$ of Refs.~%
\onlinecite{CaC6:calandra:band,CaC6:jskim:press} shows that a considerable
fraction ($60\%$) of the total e-ph coupling comes from modes which involve
carbon atoms and not the intercalant ion. This observation, together with
the empirical correlation between the filling of the IL band and
superconductivity, suggests that, in analogy with the hole-doped graphite
compound MgB$_{2}$, CaC$_{6}$ and YbC$_{6}$ might be understood as
\textquotedblleft electron-doped graphite\textquotedblright , where a
different source of e-ph interaction is switched on when electrons are
driven into the IL band.

In this paper we will show, via density-functional-theory
calculations,\cite{DFT} that indeed a strong interband e-ph
interaction between $\pi^*$ and IL electrons, due to out-of-plane
phonons, survives in graphite even when the intercalant atom is
replaced by a uniform positive background of charge
(\emph{Jellium-Intercalated-Graphite, JIG}).\cite{GIC:csanyi:band}
Such a finding contrasts with the rough idea that the IL states,
being free-electron like, cannot interact with the surrounding
lattice: this may become true when the carbon layers are very far apart from
each other, 
but not at the spacings observed in real GICs, where, as our results clearly
show, 
the IL states are strongly affected both by the presence and by the dynamics 
of the carbon layers which sandwich them.
Similar confined, free-electron states are found in several layered materials\cite{boride:medvedeva:band,CaAlSi:gianto:band,LiB:curtarolo:band,LiB:mazin:band}
 and even in nanotubes,~\cite{NFES_in_CNTs} where our results
 suggest that similar e-ph coupling interactions may take place.

The paper is organized as follows: in the first section we discuss
the electronic structure of JIG with the aid of Wannier functions.
In the second and third sections we calculate the vibrational
spectrum and e-ph interaction of JIG. We then consider these as a
function of doping and interlayer spacing in the fourth section,
comparing  the results of our model with the available
experimental data on GICs. 
Finally, in Appendix A, we provide a minimal $3\mathrm{%
\times }3$ tight-binding model for JIG, while Appendix B contains
the computational details for the different methods used: planewaves
and pseudopotentials,~\cite{PWscf} TB-LMTO~\cite{Theory:andersen:TBLMTOASA} and
NMTO.~\cite{Theory:andersen:NMTO}

\section{Electrons}

\label{sect:bands}

\begin{figure}[tbh]
\begin{center}
\includegraphics[width=6cm, angle=-90]{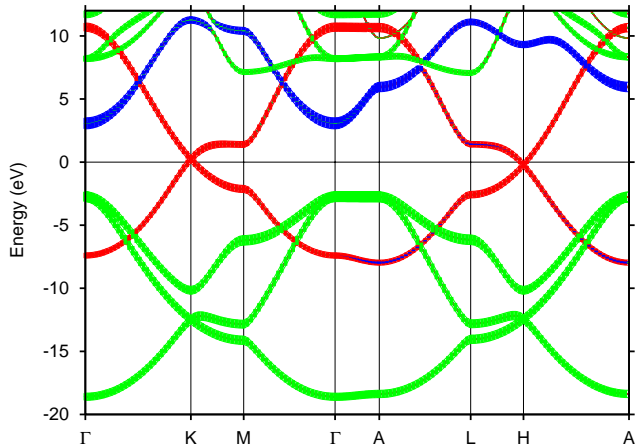}
\end{center}
\caption{(\emph{Color Online}) Band structure of primitive graphite. The
fatness (and color) of the bands is proportional to the partial contribution
of each orbital channel: green ($\protect\sigma $), red ($\protect\pi $) and
blue (interlayer $\equiv $ IL).~\protect\cite{fatbandcharacter} The lattice
constants used in this plot are those of CaC$_{6}$ (see text and note~\onlinecite{CaC6structure}). 
The Fermi level of CaC$_{6}$ would in this
simple picture be at about 4.5 eV, see also Fig.$\,$\protect\ref{fig:fig9}.}
\label{fig:fig1}
\end{figure}
A good starting point to understand the more complicated band structure
of actual GICs, such as CaC$_6$ and YbC$_6$, is represented by primitive
graphite. 
In this structure (space group 191) the carbon hexagons
sit on top of each other ($\alpha \alpha \alpha $ stacking),\textit{\ i.e.},
they are not laterally shifted like in real graphite;
the primitive cell thus contains two carbon atoms (C$_2$).

The color and fatness~\cite{fatbandcharacter} of the electronic bands 
of C$_2$, shown in 
 Fig.~\ref{fig:fig1},
indicate the relative importance of the
partial contribution of each orbital channel to the given band: green ($%
\sigma $), red ($\pi $), and blue (IL).

With $\alpha \alpha \alpha $ stacking and negligible dispersion
perpendicular to the sheets, the bonding $\pi $ and antibonding $\pi ^{\ast
} $ bands touch at a (Dirac) cone in energy-versus-momentum space; with
an occupancy of 4 electrons per C, the Fermi surface consists of merely the
K-points in the Brillouin zone; simple graphite is an ideal semimetal. In
the simplest rigid-band picture, hole-doping shifts the Fermi level down,
towards and eventually inside the $\sigma $ bands, like in MgB$_{2}.$
Electron-doping shifts the Fermi level up, and eventually brings it into the
IL band (CaC$_{6}$, YbC$_{6}$, and other GICs).

The main difference between the rigid-band picture and actual GICs is that the intercalation of positive ions between the graphene layers shifts the IL bands down in energy with respect to the $\pi $ bands, without substantially modifying their shape in the relevant energy window;\cite{GIC:csanyi:band,GICs:Mazin:review} this effect is reproduced quite well even if one replaces the intercalant ions by a uniform background of positive charge (JIG), provided that the interlayer spacing
is fixed at the corresponding value of the actual GIC.\cite{GIC:csanyi:band}

While the nature of the $\sigma $ and $\pi $ bands has been extensively discussed in literature, less information is available about the IL band in GICs.
IL states have mainly been described as three-dimensional, nearly-free electrons in the empty interlayer region, whence their name; and indeed, as far as the motion parallel to the graphite planes is concerned (the $xy$ directions), the IL electrons, to a very good approximation, freely propagate in the interstitial volume, with plane-wave-like wavefunction ${\rm exp}(ik_x x+ik_y y)$ and energy dispersion $\frac{1}{2}(k_x^2+k_y^2)$.

In the $z$ direction, however, their motion is not free, being periodically damped, with period $c$, by the graphene layers, across which the IL electrons are only partially transmitted. Along $z$, these layers act like the periodic potential barriers of the Kronig-Penney model,~\cite{Kittel} where the free-electron parabola $\frac{1}{2}k_z^2$ is split into a set of sub-bands separated by gaps, with sub-bandwidths and gaps dictated by the strength of barriers and by their spatial separation $c$. In this model all gaps close, and the full 3D free-electron dispersion is recovered, for vanishing barrier strength; at fixed $c$, as the barrier strength increases, the transmission across potential barriers decreases, until, for infinite strength, electrons are  confined between pairs of barriers, yielding a set of completely flat sub-bands; in our case this translates into IL bands $\varepsilon_m = \frac{1}{2}(k_x^2+k_y^2)+ \frac{1}{2}(m\pi/c)^2, m=1,2,...$ with no dispersion along $k_z$. 

A careful examination of different GICs (and of C$_2$ for several values of $c/a$) confirms that their IL bands stay somewhere between these two limiting cases, with the graphene layers approximately behaving, along $z$, as an array of barriers with fixed finite strength and variable period $c$, which depends on the intercalant atom. These bands have a substantial dispersion along $k_z$ --so there is significant transmission across the graphene layers, or, in other words, the IL electrons are not completely confined; but their slope and width strongly depend on $c$ --so the partial reflection against the graphene layers is  significant too.

In the range of parameters analyzed in the present  paper, which correspond to existing donor-intercalated graphites, the lowest $m=1$ Kronig-Penney band is well separated in energy from the others --another evidence that, in the $z$ direction, the nearly-free-electron theory does not hold for the IL states-- and it is the only one which can be occupied; to this band we refer, in the following, as {\em the} IL band (blue in Fig.$\,$\ref{fig:fig1}). 
\begin{figure}[tbh]
\begin{center}
\includegraphics[width=8.5cm]{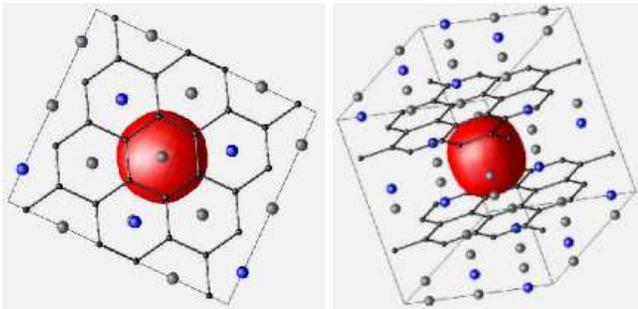}
\end{center}
\caption{(\emph{Color Online}) Top and side view of the 
Wannier-like~\cite{NMTOAppA} function $\protect
\psi _{s}(\mathbf{r})$, see also Appendix A, for the IL band (blue in Fig.\ref{fig:fig1}).
The basis set spanning the IL band consists of the replicas of
this function placed at all interstitial sites (grey and blue spheres). The
figure shows the contour $\left\vert \protect\psi _{s}(\mathbf{r})
\right\vert =0.07\times (a.u.)^{-3}$, with $\protect\int \protect\psi _{s}^{2}d^{3}r=1.$ The sign of $\protect\psi _{s}(\mathbf{r})$ is indicated by red and green.}
\label{fig:fig2}
\end{figure}

In Fig.~\ref{fig:fig2} we show the shape of its  Wannier function\cite{NMTOnote}. 
Because the IL band overlaps other bands
--the $\pi ^{\ast }$ and $\sigma ^{\ast }$ bands, with which it,
however, hybridizes very little,-- and because the function shown
in Fig.~\ref{fig:fig2} has not yet been symmetrically orthogonalized
to the lattice-translated functions, it ought to be called
Wannier-like as in Refs. \onlinecite{Theory:andersen:NMTO} and
\onlinecite{ZurekNMTO}. The basis set spanning the IL band,
interpolated smoothly across any hybridization gap, consists of this
Wannier-like function and its replicas displaced by all lattice
translations. We have chosen the IL Wannier-like function to be
centered on an interstitial site, where it is seen to have the full
point-group symmetry and a simple, $s$-like shape. The bond-centered
Wannier-like function for the $\pi ^{\ast }$ band,\cite{ZurekNMTO} interpolated
smoothly across hybridization gaps, is the antibonding linear
combination of the two C$\,p_{z}$ orbitals at the vertices of the
bond, plus a halo of further antibonding $C$ $p_{z}$ orbitals. The
basis set spanning the $\pi ^{\ast }$ band consists of this
Wannier-like function and its replicas displaced by all lattice
translations. That is, every second C-C bond carries such a
function. Since the IL Wannier-like function has everywhere the same
sign, while the sign of the $\pi ^{\ast }$ Wannier-like function
alternates within the scale of the IL function (there is one IL
function for every two C $p_{z}$ orbitals), the two functions, one
displaced against the other by an arbitrary lattice translation, are
very nearly orthogonal. This is the reason why the IL and $\pi
^{\ast }$ bands hardly hybridize, as seen in Fig.~\ref{fig:fig1}. On
the contrary, the Wannier-like function for the $\pi $ band\cite{ZurekNMTO} is the
bonding linear combination of the two C$\,p_{z}$ orbitals at the vertices of the
bond, and that is not orthogonal to the IL function. For that reason, the $\pi
$ band may hybridize with the IL band, as will be discussed in
Appendix A. This hybridization shows up if, by doping or
changes in the $c/a$ ratio, the IL band moves into the energy region of the $%
\pi $ band.

Based on the band and orbital picture of primitive graphite
(C$_2$) discussed  so far, in the following sections we will 
study the electron-phonon properties of JIG, {\em i.e.}
an artificially electron-doped C$_2$, where 
the electrons in excess are neutralized  by
 a uniform, positive  background charge;
the comparison of the band structure of JIG with that
of real GICs has shown that 
two important independent variables
govern the filling of the interlayer band:
 the interlayer spacing $c$ and
the number of electrons added to the system. In our study,
these two parameters are represented by
the doping $n$, defined as the number of electrons added to C$_{2},$ and
neutralized by a jellium background of the same magnitude
(the number of C valence electrons is thus $4+n/2$) and the $c/a$ ratio
of a C$_{2}$ unit cell with the same in-plane lattice constant as CaC$_6$.
To minimize the number of parameters we in fact kept the in-plane lattice
constant fixed throughout our study, so varing $c$ is like varying $c/a$.
\section{Phonons}

\label{sect:phonons}

The top left part of Fig.~\ref{fig:fig3} shows the phonon dispersions
we calculated for the doping levels $n\mathrm{=}0$ (dotted, pure graphite) and $n\mathrm{=}2/3$ (corresponding to CaC$_{6}$, full line). 
We used the interlayer spacing $c$ and planar lattice constant $a$ of CaC$_{6}$ in
both cases, to compare with available experimental and theoretical data. With two atoms per cell there are 6 phonon branches, and since
a phonon displacement is a vector like an electronic C$\,p$ function, the 6
phonon branches are topologically equivalent with the electronic C $\,p$
bands, that is, with the $\pi $ and $\pi ^{\ast }$ bands and with the non-$s$
part of the $\sigma $ and $\sigma ^{\ast }$ bands in Fig.$\,$\ref{fig:fig1}.
In particular, the two phonon branches which are the softest along KH, are
the out-of-plane vibrations with dispersions similar to those of the $\pi $
bands in Fig.$\,$\ref{fig:fig1}; the four other branches are the in-plane
vibrations. In-plane and out-of-plane modes hardly mix. For that reason, the
$n\mathrm{=}0$ and $n\mathrm{=}2/3$ phonon dispersions are qualitatively
similar.

The only qualitative difference occurs along the $\Gamma $-A-L path, where
the out-of-plane acoustic branch of JIG has imaginary frequencies.
Such an instability is an artifact of JIG: at $n=0$
this branch is just very soft and sensitive to computational 
details, but not unstable;\cite{graf:marzari:band}
in actual GICS the intercalants contribute to the inter-layer binding and
stabilize the strucure. Fortunately this artificial JIG instability
occurs where the e-ph interaction is unimportant 
(no green dots where the frequencies are negative in the upper left panel of
Fig.~\ref{fig:fig3}), and does not affect our subsequent analysis.

In real CaC$_{6}$ and other GICs~\cite{GIC:calandra:band} 
in- and out-of-plane C modes are also separated in energy, 
 and occur at
similar energy scales, and they do not mix with the intercalant modes, which
are at a lower energy scale ($\omega \sim 10$ meV in CaC$_{6}$~\cite%
{CaC6:calandra:band, CaC6:jskim:press}).

\section{Electron-Phonon coupling}
\subsection{Jellium Intercalated Graphite (JIG)}
In a metal, e.g. for doped graphite, the $\nu \mathbf{q}$ phonon can couple
between electrons $n\mathbf{k}$ and $m\left( \mathbf{k+q}\right) $ at the
Fermi surface, and thereby attain a relative line-width:%
\begin{equation}
\frac{\gamma _{\nu \mathbf{q}}}{\omega _{\nu \mathbf{q}}}=2\pi \sum_{nm%
\mathbf{k}}|g_{\nu ,\,n\mathbf{k},\,m\left( \mathbf{k+q}\right) }|^{2}\delta
(\varepsilon _{n\mathbf{k}})\delta (\varepsilon _{m\mathbf{k+q}}).
\label{linewidth}
\end{equation}%
Here, $\sum_{\mathbf{k}}$ is the average over the Brillouin zone and%
\begin{equation}
g_{\nu ,n\mathbf{k},m\left( \mathbf{k+q}\right) }=\left\langle n\mathbf{k}%
\left\vert \delta V\left( \mathbf{r}\right) \right\vert m\left( \mathbf{k+q}%
\right) \right\rangle /\delta Q_{\nu \mathbf{q}}  \label{g}
\end{equation}%
is the matrix element of the perturbation, $\delta V\left(\mathbf{r}\right)
,$ of the self-consistent electronic potential due to the displacement,%
\begin{equation*}
\delta R_{\alpha j}=\frac{e_{\alpha j,\nu \mathbf{q}}}{\sqrt{2M_{j}\omega
_{\nu \mathbf{q}}}}\delta Q_{\nu \mathbf{q}},
\end{equation*}%
in the $\alpha $-direction of the $j$th atom (and continued as a wave with
wavevector $\mathbf{q),}$  $e_{\alpha j,\nu \mathbf{q}}$ being the
eigenvector of the $\nu \mathbf{q}$ phonon.

The squared e-ph matrix element is usually a smoother function of $\mathbf{k}
$ and $\mathbf{k+q}$ than $\delta (\varepsilon _{n\mathbf{k}})\delta
(\varepsilon _{m\mathbf{k+q}}).$ In this case, it is useful to discuss the
so-called nesting functions,%
\begin{eqnarray}
\chi _{nm}(\mathbf{q}) &\equiv &\sum_{\mathbf{k}}\delta (\varepsilon _{n%
\mathbf{k}})\delta (\varepsilon _{m\,\mathbf{k+q}})  \label{eq:nesting} \\
&=&\int_{\varepsilon _{n\mathbf{k}}=\varepsilon _{m\left( \mathbf{k+q}%
\right) }=0}\frac{1}{\left\vert \mathbf{v}_{n\mathbf{k}}\times \mathbf{v}%
_{m\left( \mathbf{k+q}\right) }\right\vert }\frac{dk}{\Omega_{BZ}}  \notag \\
&=&-\lim_{\omega \rightarrow 0}\frac{1}{\pi \omega }\Im\chi
_{nm}^{0}\left( \mathbf{q},\omega \right) .  \notag
\end{eqnarray}
which may be expressed as line integrals\textbf{\ }(in \emph{one}
Brillouin-zone, of volume $\Omega_{BZ}$) along the cut of the $n$ and the $m$ sheets
of the Fermi surface, with the latter being displaced by $-\mathbf{q}.$
Here, $\mathbf{v}_{n\mathbf{k}}\equiv \delta \varepsilon _{n\mathbf{k}%
}/\delta \mathbf{k}$ is the Fermi velocity. This is the second line of Eq.~\ref{eq:nesting}.
Its last line relates the nesting functions to the bare
susceptibilities. Note that the sum over all $\mathbf{q}$ simply yields
the product of the $n$ and $m$-sheet state densities per spin at the Fermi
level $\left( \epsilon _{F}\mathrm{\equiv }0\right) $:%
\begin{equation}
\sum_{\mathbf{q}}\chi _{nm}(\mathbf{q})=\sum_{\mathbf{kk}^{\prime }}\delta
(\varepsilon _{n\mathbf{k}})\delta (\varepsilon _{m\,\mathbf{k}^{\prime
}})=N_{n}\left( 0\right) N_{m}\left( 0\right) ,  \label{DOS}
\end{equation}%
because $N_{n}\left( 0\right) \equiv \sum_{\mathbf{k}}\delta (\varepsilon _{n%
\mathbf{k}}).$
Finally, the probability for an electron to scatter from and to any point of
the Fermi surface with total density of states $N\left( 0\right) \equiv
\sum_{n}N_{n}\left( 0\right) $, due to the interaction with a phonon of
frequency $\omega $, is given by the Eliashberg spectral function%
\begin{gather}
\alpha ^{2}F(\omega )=  \label{Eliashberg} \\
\frac{1}{N\left( 0\right) }\sum_{nm\mathbf{k}}\delta (\varepsilon _{n\mathbf{%
k}})\delta (\varepsilon _{m\mathbf{k+q}})\sum_{\nu \mathbf{q}}|g_{\nu ,\,n%
\mathbf{k},\,m\left( \mathbf{k+q}\right) }|^{2}\delta (\omega -\omega _{\nu
\mathbf{q}}).  \notag
\end{gather}%
The total e-ph coupling constant is then:%
\begin{equation}
\lambda =2\int_{0}^{\infty }\frac{\alpha ^{2}F(\omega )}{\omega }d\omega
=\sum_{\nu \mathbf{q}}\lambda _{\nu \mathbf{q}},  \label{eq:lambda}
\end{equation}%
and the last equation uses Eq.~\ref{linewidth} for the phonon
line-widths to express $\lambda $ as a sum over mode coupling constants%
\begin{equation}
\lambda _{\nu \mathbf{q}}\equiv \frac{1}{\pi N\left( 0\right) }\frac{\gamma
_{\nu \mathbf{q}}}{\omega _{\nu \mathbf{q}}^{2}}.  \label{eq:lambda_qnu}
\end{equation}

\begin{figure}[t]
\begin{center}
\includegraphics*[width=8cm]{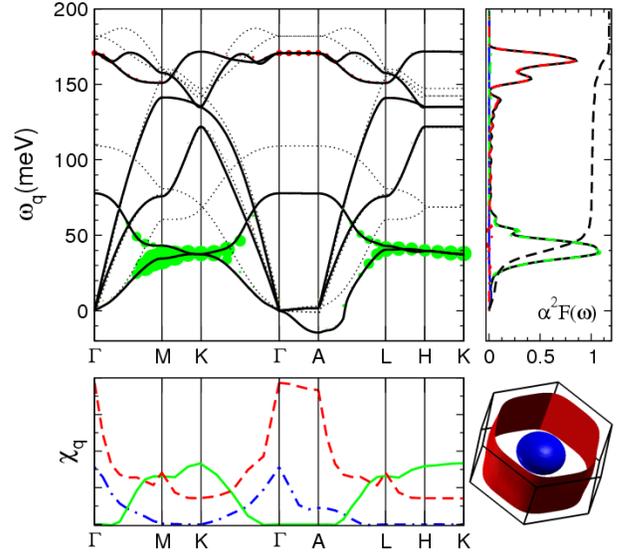}
\caption{\label{fig:fig3}
(\emph{Color Online}) \emph{Upper left}: phonon dispersions
(black solid lines) for JIG with the same lattice constants $\left(
c/a = \text{1.81}\right) $ and filling $n\!\!=$2/3 as in CaC$_{6}$.
Thin dotted lines show the corresponding dispersions for $n\!=$0
(undoped graphite). Imaginary frequencies are shown as negative (see
text). The area of a
colored dot is proportional to the mode coupling constant $\protect\lambda _{%
\protect\nu \mathbf{q}}.$ Color code: red $\protect\pi ^{\ast }$ intraband;
blue IL intraband (negligible); green $\mathrm{IL}\!\mathrm{-\!}\protect\pi %
^{\ast }$ interband. \emph{Upper right}: JIG e-ph spectral function $%
\protect\alpha ^{2}F(\protect\omega )$ and $\protect\lambda (\protect\omega %
) $ defined as in Eq.~\protect\ref{eq:lambda}, but integrated to $\protect%
\omega $ rather than to $\infty $ (dashed). \emph{Lower left}: JIG
nesting functions $\protect\chi _{nm}(\mathbf{q}).$ \emph{Lower
right}:
Fermi surface of JIG in the first Brillouin zone (IL-sphere and $\protect%
\pi ^{\ast }$-cylinder).}
\end{center}
\end{figure}

These quantities are shown for JIG in Fig.~\ref{fig:fig3}; in the
upper half we focus on the vibrational properties and the e-ph
coupling (phonon dispersions and partial Eliashberg functions) and
in the lower half on the electronic properties 
(nesting functions and Fermi surface).

We see that the Eliashberg function $\alpha ^{2}F(\omega )$ merely has \emph{%
two} pronunced peaks, respectively centered at $\omega \mathrm{\sim }40\,$%
meV and $\omega \mathrm{\sim }160\,$meV. The first involves the carbon
out-of-plane, zone-boundary modes, and the second involves the in-plane
bond-stretching, zone-center modes. The same two peaks (compare our Fig.~\ref%
{fig:fig3} with Fig.~3 of Ref.~\onlinecite{CaC6:calandra:band}) are found in
the $\alpha ^{2}F(\omega )$ of CaC$_{6}$, which, of course, also has an
additional low-frequency peak involving in-plane vibrations of Ca.

Integrating $\alpha ^{2}F(\omega )/\omega $ over positive frequencies~\cite{lambdaNOTE}
yields: $\lambda \mathrm{=}1.1$, thus even larger than in CaC$%
_{6},$ where $\lambda \mathrm{=}0.84,$ despite the absence of
the calcium modes. 
We have thus shown that a large e-ph
interaction survives in JIG even after removing the intercalants,
whereas all the recent studies on superconductivity in GICs have
 assumed that either the  mechanism
for  superconductivity
is not conventional e-ph coupling,~\cite{GIC:csanyi:band}
 or that it is
intrinsically related to the presence of the intercalant.~\cite%
{GIC:mazin:band,CaC6:calandra:band}
The discussion of possible reasons for the quantitative
discrepancy between the $\lambda's$ of JIG and CaC$_6$ is postponed
to the end of the section.

The Fermi surface of JIG (bottom right panel of Fig.~\ref{fig:fig3})
has two sheets, both of
electron character:
the sum of their volumes is $n\mathrm{\times }%
\Omega_{BZ}/2=\Omega_{BZ}/3.$ The $\pi ^{\ast }$ sheet has negligible $k_{z}$-dispersion and
is open, with the shape of honeycomb enclosing the vertical KM-HL BZ
boundaries. Its hollow inside forms a $\Gamma $A-centered cylinder with
average radius $k_{\pi ^{\ast }}$ as shown in red at the lower right of Fig.~%
\ref{fig:fig3}. The IL sheet, shown in blue, is three dimensional, $\Gamma $%
-centered, and closed; when $n$ and $c/a$ equal those of CaC$_6$, 
as in Fig.~\ref{fig:fig3}, it closely approximates a sphere with radius $k_{\mathrm{%
IL}}$. 

The e-ph interaction  can take place in three channels: IL
intraband, $\pi ^{\ast }$ intraband, and IL-$\pi ^{\ast }$
interband. The $\lambda _{\nu \mathbf{q}}$'s associated with each
phonon mode (Eq.~\ref{eq:lambda_qnu}) are shown in the upper left of
Fig.$\,$\ref{fig:fig3} by full dots with area proportional to
$\lambda _{\nu \mathbf{q}}.$ 
The contributions from the IL and $\pi ^{\ast }$ intraband scattering 
are respectively blue and red, consistently with the color code used
for these two bands in Fig.~\ref{fig:fig1} and for the corresponding Fermi surface sheets in the lower right panel of Fig.~\ref{fig:fig3}.
The interband IL-$\pi ^{\ast }$ scattering contribution, is, instead,
green. The upper right of Fig.$\,$\ref{fig:fig3} shows the
Eliashberg function and the lower left the nesting functions with
the same decomposition scheme and color code.

Fig.~\ref{fig:fig3} clearly shows that the zone-boundary out-of-plane
buckling phonons with $\omega \mathrm{\sim }40\,$meV give the largest
contribution to the total $\lambda $, and that the coupling is due to
interband IL-$\pi ^{\ast }$ coupling. 
The $\pi ^{\ast }$ intraband coupling by the optical
in-plane bond-stretching phonons~\cite{C:piscanec:band} cause the large $%
\alpha ^{2}F(\omega )$ peak at $\omega \mathrm{\sim }160\,$meV. Due to the
high frequency, however, this peak contributes a mere 20\% to the total $%
\lambda $. Finally, with the absence of intercalant modes, our JIG
model has no phonons with small $\left\vert \mathbf{q}\right\vert $
causing significant coupling  (IL intraband and interband).

To understand the origin of the large interband e-ph matrix element in Eq.~\ref{g}, we go back to the Wannier-like orbital in Fig.~\ref{fig:fig2}. As
we saw, the IL and $\pi ^{\ast }$ bands cross essentially without hybridizing (a
tiny hybridization occurs only along the A-L line) because the IL
Wannier-like orbital is invariant under all operations of the point group of
the interstitial site, while the sign of the Wannier-like function for the $%
\pi ^{\ast }$ band alternates within the extent of the IL function. If,
however, we apply a buckling distortion to the graphene sheets, half of the $%
p_{z}$ orbitals are driven closer to the interstitial site where the
IL orbital of Fig.~\ref{fig:fig2} sits,
and the other half are driven away, with the net result of a non-zero e-ph
matrix element between the $\pi ^{\ast }$ and IL bands (see the Appendix A,
where this is demonstrated using a TB model).

\begin{figure}[tbh]
\begin{center}
\includegraphics[width=6.0cm]{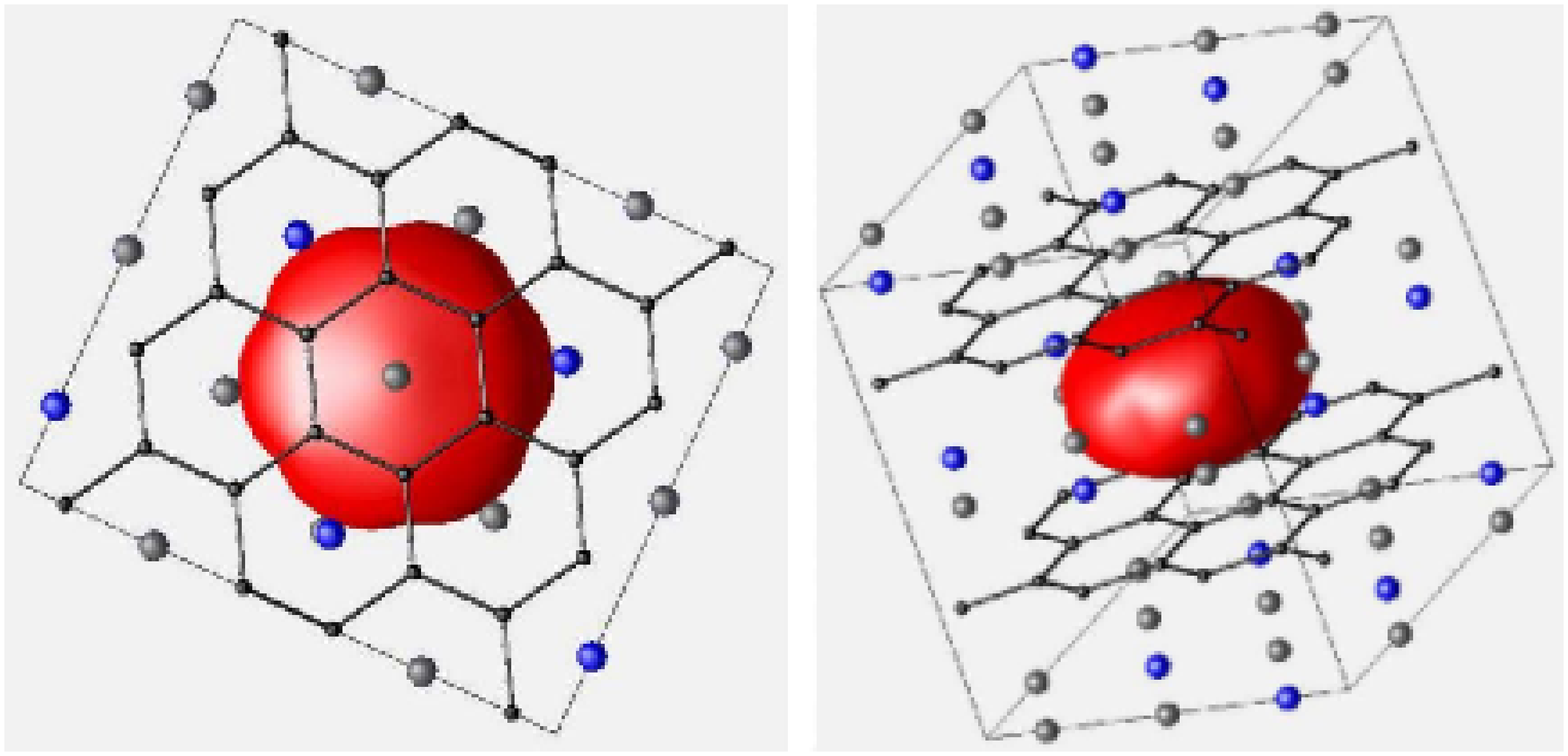}
\par
\includegraphics[width=6.0cm]{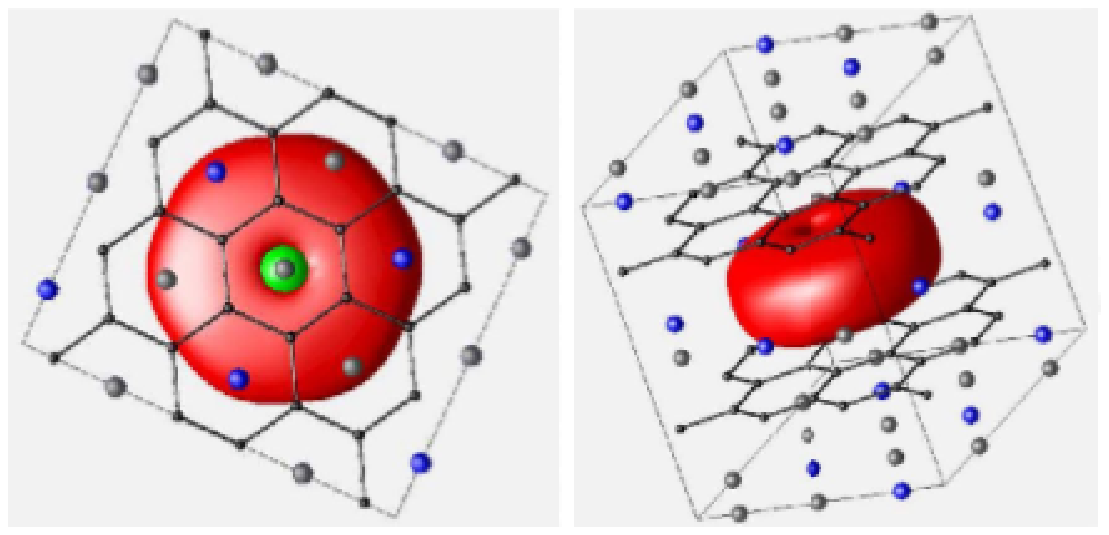}
\end{center}
\caption{(\emph{Color Online}) Top and side view of the Wannier-like
function for the lowest of the three IL subbands, the $a_{1g}$
subband, in JIG (top) and in CaC$_{6}$ (bottom). The basis set
spanning the $a_{1g}$ subband consists of the replicas of this
function placed at all Ca sites, i.e., at every third interstitial
site, the blue ones. The isocontour and the color scheme is the same
as in Fig. \protect\ref{fig:fig2}.} \label{fig:all-NMTO}
\end{figure}

In fact, we only need the e-ph matrix elements for Bloch functions at the
Fermi level, and not for the entire IL band. Since the $n$=2/3 doped
electrons not only go into the IL band, but also, and even more, into the $%
\pi ^{\ast }$ band, the former is less than 1/3 full. To calculate the IL
Bloch functions at $\varepsilon _{F}$ we could therefore also use the
Wannier-like function for the lowest IL subband, obtained by folding the IL
band into e.g. the three times smaller BZ of CaC$_{6}.$ This function is
shown in the upper part of Fig.~\ref{fig:all-NMTO} using the same contour as
in Fig.~\ref{fig:fig2}. It is seen also to have the full point-group
symmetry, and since the basis set spanning the lowest IL subband consists of
the replicas of this function placed at all Ca sites, that is, at every
third interstitial site, the square of this function is three times more
diluted (extended) than the square of the Wannier-like function for the
entire IL band. Whereas the Wannier function for the lowest IL subband has $%
a_{1g}$ symmetry, the ones for the two upper IL subbands have $e_{g}$ symmetry,
that is, their signs alternate. So one could also say that, for these three subbands, the Wannier function can be obtained by forming the three linear combinations (of  axial symmetry $a_{1g}$ and $e_{g}$) of the ``small'' Wannier function of the entire IL band (Fig.~\ref{fig:fig2}). For the lowest band this yields the ``large'' Wannier function shown in Fig,~\ref{fig:all-NMTO}, whose axial symmetry is $a_{1g}$. The point is, that for the relatively low fillings of the IL band which are of our interest, this is the Wannier function required to form the proper Bloch IL functions at the Fermi
level; and this function \emph{\ is uniform on the scale of the C-C bond length.} For that reason there is a large, "dormant" IL-$\pi ^{\ast }$ matrix element for
buckling modes.

\subsection{JIG vs. CaC$_6$}

How well do our results for JIG reproduce what is observed in real GICs? 
As an example, we consider CaC$_6$. 
Here, the IL band is folded three times into the rhombohedral BZ, and split;
however, the occupied band structure of CaC$_6$ still bares a very close
resemblance to that of JIG.~\cite%
{GIC:csanyi:band,GIC:mazin:band, CaC6:calandra:band,GICs:Mazin:review}
 This is clearly visible in Fig.~\ref{fig:fig7}, where we show an in-plane cut of the Fermi surface of the two systems in the  (three times smaller) rhombohedral
Brillouin zone of CaC$_{6}$.~\cite{CaC6structure}
The folding of the hexagonal BZ of primitive graphite into the rhombohedral BZ of CaC$_{6}$, shown in the left panel, may obscure the simple Fermi surface topology appearing in the lower right panel of Fig.~\ref{fig:fig3}, but it clarifies the origin of the complicated Fermi surface of CaC$_{6}$, shown in the right panel of Fig.~\ref{fig:fig7}. In both cases we clearly recognize a quasi-spherical IL Fermi surface
and a more complicated structure, which, however, only amounts to folding
the outer  $\pi^{\ast}$ honeycombs;  in the right panel only small hybridization gaps,
in the $\Gamma-M$ and equivalent directions, make CaC$_6$ slightly different from the purely folded graphite shown at the left.

\begin{figure}[tbp]
\caption{(\emph{Color Online}) Fermi surface of JIG and of of
CaC$_{6}$, both shown in the reciprocal space of rhombohedral
CaC$_{6}$ (top view, compare to the 3D view of
Fig.~\protect\ref{fig:fig3}, lower right panel).
Once folded into the three times smaller BZ, the JIG Fermi surface (\emph{%
left}), which in the hexagonal BZ (grey dotted lines) consists of an IL
sphere (blue) and a $\protect\pi ^{\ast }$ honeycomb (red), reveals
remarkable similarities with that of CaC$_{6}$ (\emph{right}). }
\label{fig:fig5}
\includegraphics*[width=8cm]{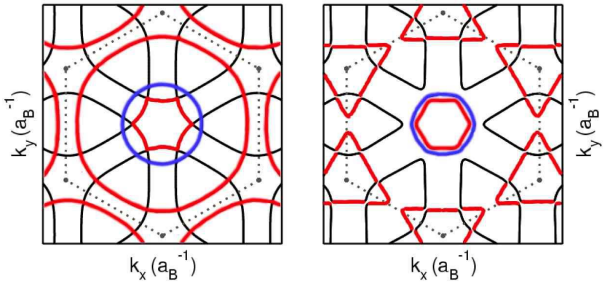}
\end{figure}
This comparison suggests that the introduction of Ca atoms between the
C$_2$ layers preserves the physical picture discussed above.
This is also confirmed by the  Wannier-like function for the
lowest IL subband, shown in the bottom panels of Fig.~\ref{fig:all-NMTO},
and is seen to be very similar to the one for the lowest subband of JIG,
shown in the top panels and discussed before.
The main difference is the presence of some Ca$\,$3$d_{3z^{2}-1}$ character
in the IL Wannier-like function for CaC$_{6}$, but this does not disrupt 
the large dormant IL-$\pi^{\ast}$ matrix element for buckling modes. 
\subsection{Why JIG overestimates the CaC$_6$ e-ph}
A large e-ph interaction associated to C out-of-plane
modes has been calculated in CaC$_6$ and several 
other GICs;~\cite{CaC6:calandra:band,CaC6:jskim:press,GIC:calandra:band} 
our JIG model gives a convincing 
qualitative picture of this electron-phonon interaction in CaC$_6$ and,
as shown in the next section, also correctly explains the
observed trend of $\lambda$ observed in different GICs.
However, as we already pointed out, there is a large quantitative 
discrepancy in the values of $\lambda $, even though the electronic 
parameters (DOS, Fermi velocities) are comparable.
As a matter of fact, in JIG we obtain an overall larger $\lambda$ than in the corresponding real GICs, even though, in JIG, the phonon modes associated to the in-plane vibrations of the intercalant, 
which give a sizable contribution to $\lambda$ in real intercalated graphites
CaC$_{6}$, SrCa$_{6}$, and BaC$_{6}$, are missing (these calcium 
modes cause a strong e-ph interaction, because displacing the IL 
Wannier-like function off center in the $xy$-plane also awakes the 
dormant IL-$\pi ^{\ast }$ and IL-IL couplings). 

The bottomline is that, in JIG, the $\lambda$ associated to the IL-$\pi ^{\ast }$
coupling via out-of-plane phonons is almost three times as large as in 
real GICs. Why? Most of this boost comes from the difference in phonon 
frequency: in real compounds, the presence of an intercalant
ion partially hinders the vertical motion of C atoms, thus hardening
the out-of-plane frequencies.

For example, the peak associated to C out-of-plane vibrations is at $\omega
\mathrm{\sim }$55 meV in CaC$_{6}$, but at $\omega \mathrm{\sim }$40
meV in the corresponding JIG. Assuming that the partial $\lambda $
associated to this mode can be approximated by the Hopfield formula
$\lambda =N(0)g^{2}/M\omega ^{2}$, where $g$ is the e-ph matrix element,
 this would give an enhancement factor of  $(55/40)^{2}$, i.e., almost two,
going from real CaC$_6$ to JIG, if $N(0)$ and $g$ were
the same in the two systems. 
But they are not, and the quantitative differences in the band
structure which affect them (such as a weak hybridization between IL and $%
\pi ^{\ast }$ bands, visible in CaC$_{6}$ but absent in JIG, see Fig.~\ref%
{fig:fig7}) are probably enough to explain the total enhancement
factor of almost 3 found between JIG and actual CaC$_{6}$, as far as
the e-ph coupling of carbon-buckling modes is concerned.


\section{Effects of electron doping, $n,$ and interlayer spacing, $c.$}
\label{sect:doping-spacing} 
Csanyi et al.~\cite{GIC:csanyi:band}
have proposed that the electronic properties of donor-intercalated 
GICs be mainly determined by the interlayer spacing  and
valence of the intercalant; recent first-principles 
calculations~\cite{GIC:calandra:band} and experiments~\cite{GIC:jskim:unp}
suggest a possible
correlation between T$_c$ and the $c$ axis lattice constant;
 these hints are both confirmed and clarified in their origin
   by the results presented here.
Having so far studied in detail the electron-phonon
interaction of JIG in comparison with CaC$_6$,
we will now turn to discuss how the two independent variables of
our model, $c/a$ and $n$, affect the e-ph properties of JIG, and 
compare the JIG trends with the available experimental data.
In the whole range of $c/a$
and $n$ considered here, there is a single interlayer band at the
Fermi level (see Section I ).

\begin{figure}[t]
\begin{center}
\includegraphics*[width=8.5cm]{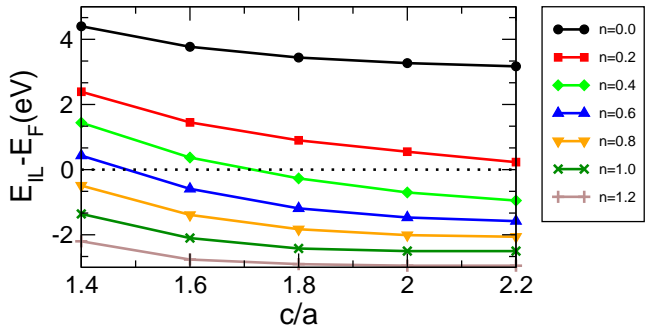}
\end{center}
\caption{(\emph{Color Online}) Position of bottom the IL band in JIG
 as a function
of interlayer C spacing ($c/a$) for different electron dopings $n$;
see also Fig.4 of Ref.~\onlinecite{GIC:csanyi:band}.}
\label{fig:fig6}
\end{figure}

 \emph{Electrons:} 
In Fig.~\ref{fig:fig6} we plot the energy position of the bottom of the
IL band as a function of $c/a$ for different $n$.
As mentioned in Section I,
the addition of a uniform, positive
background of charge to graphite shifts the IL band down in energy
w.r.t. the rest of the band structure, leaving its shape unaffected.  
The effect is obviously stronger for larger $n$.
The IL band is shifted down in energy also when the $c/a$ ratio is
increased. Changing the separation between the graphite planes 
also sensibly affects the IL bandwidth along $k_z$, 
so that the associated FS, which in CaC$_6$ is nearly 
a sphere (Fig.~\ref{fig:fig3}),
becomes an oblate ellipsoid for smaller $c$'s. In the opposite limit
(large $c$'s), the opposite happens.
The same trend is observed 
in real GICs (see Ref.~\onlinecite{GIC:calandra:band}).

\emph{Phonons:} In the upper left panel of Fig.~\ref{fig:fig3} we
show the phonon dispersions with electron doping $n$=2/3
(black solid line) and without doping ($n$=0, thin dotted line). 
The JIG phonon frequencies are lower than those of undoped graphite
in the entire BZ, regardless of the presence or absence of e-ph
coupling. Such a large, dispersionless energy shift, common to all
phonons, is not due to e-ph coupling, but rather to a softening of
all bonds, related to the increased filling of the $\pi ^{\ast } $
antibonding states.~\cite{softeningpaper} Fig.~\ref{fig:fig3} shows
that on top of this global phonon softening, there is a further $\mathbf{q}$%
-dependent softening caused by e-ph interaction.

\begin{figure}[t]
\begin{center}
\includegraphics*[width=8.5cm]{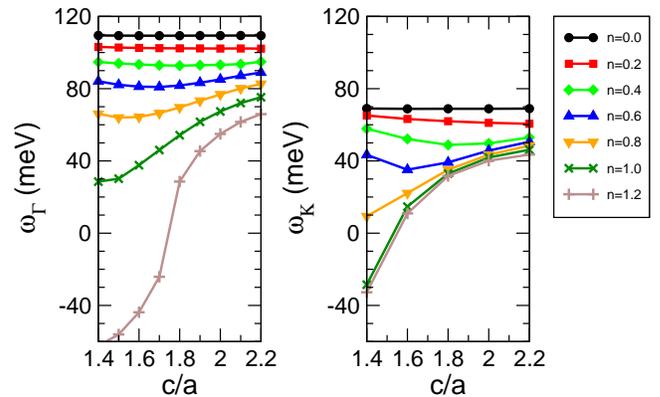}
\end{center}
\caption{(\emph{Color Online}) Frequency of the optical buckling mode at $%
\Gamma $ (left) and K (right) as a function of the $c/a$ ratio for JIG, at different dopings $n$%
; the in-plane lattice constant is that of CaC$_{6},$ for which $c/a$=1.81
and $n$=2/3$.$ Symbols represent calculated data, lines are meant to guide
the eye. Imaginary frequencies are shown as negative.}
\label{fig:fig7}
\end{figure}

The effect of doping and the $c/a$ ratio on the frequency of the
zone-center $\left( \Gamma \right) $ optical
buckling modes is illustrated in the left panel of Fig.~%
\ref{fig:fig7}. For fixed $c,$ the mode softens
for larger $n$, because of the filling of the antibonding $\pi ^{\ast }$ band.
For fixed $n$, the frequency softens at lower $c/a$, where the charge
transfer to  $\pi ^{\ast }$ states increases, due to the emptying of
the IL band (Fig.~\ref{fig:fig6}).
In the right panel of the same figure we plot the frequency of the
doubly-degenerate  buckling mode at the $\left( K \right)$ point;
here, on top of the softening due to charge transfer, 
an additional softening due to e-ph interaction takes place.

The same softening of the out-of-plane phonons with respect to pure
graphite is observed also in real GICs, where the effect is partially 
compensated by the presence of the intercalant, which hinders
the vertical motion of the C atoms.
However, in some cases the effect depicted in Fig.~\ref{fig:fig7}
is so strong as to lead to structural instabilities also 
in actual compounds. 
An extreme example in this sense is LiC$_{2}$,
for which, using  the lattice parameters of Ref.~\onlinecite{GIC:lang:theory},
we found imaginary frequencies in the \emph{entire} BZ for both the
acoustical out-of-plane and the optical buckling branch.
Fig.~\ref{fig:fig7} further suggests that many compounds with chemical formula
AC$_2$ may be dynamically unstable for this mechanism.

\emph{Electron-phonon coupling:} 
We have just seen that the largest source
of e-ph coupling in JIG is the interband IL-$\pi ^{\ast }$ channel, and
that $c/a$ and $n$ critically affect the relevant electronic bands and
phonon modes.
To understand how this can affect the e-ph properties, 
we observe that, in general, $\lambda$ 
 will benefit from three factors: 
the filling of the IL band (Fig.~\ref{fig:fig6}),
the softening of the out-of-plane phonon frequencies (Fig.~\ref{fig:fig7}),
and an increase of the IL-$\pi^{\ast}$
electron-phonon matrix element ($g$), which grows as the graphene layers are
squeezed together (not shown).
This suggests that a promising route to increase T$_c$ is either to
increase the electron concentration - but this route is severely limited
by lattice instabilities - or to reduce the $c/a$ ratio, by applying pressure
or intercalating smaller atoms.
Notice however that even in the  second route $\lambda$ cannot be increased beyond a 
threshold value, because as $c/a$ is reduced the IL band is gradually
emptied (see Fig.~\ref{fig:fig6}).
Experiments have indeed shown that pressure can increase T$_c$ in these compounds; but this
route is unfortunately limited by the appearance of instabilities in the vibrational 
spectrum.~\cite{CaC6:gauzzi:press,CaC6:jskim:press,GIC:calandra:band}

\begin{figure}[tbh]
\caption{(\emph{Color Online}) JIG Eliashberg function $\protect\alpha %
^{2}F(\protect\omega )$ (solid lines) and frequency-dependent $\protect%
\lambda (\protect\omega )$ (Eq.~\protect\ref{eq:lambda}, but only up to $%
\protect\omega$: dashed lines) for fixed $n\!$=\negthinspace 2/3 and
different $c$ and with $a$ fixed at its value in CaC$_{6}.$ For $c/a\!\leq $%
1.35 the IL band is empty and the low-frequency peak associated with the
out-of-plane C vibrations is absent. As $c/a$ grows from its value (1.81) in
CaC$_{6}$ to the one  in BaC$_{6}$ (2.11), this
peak shifts to higher frequency and its weight dramatically drops, while the
high-frequency peak of the in-plane modes, remains unchanged.
Reducing $c/a$ to $1.6$, instead, causes an increase in $\lambda$.}
\label{fig:fig8}
\begin{center}
\includegraphics*[width=8.65cm]{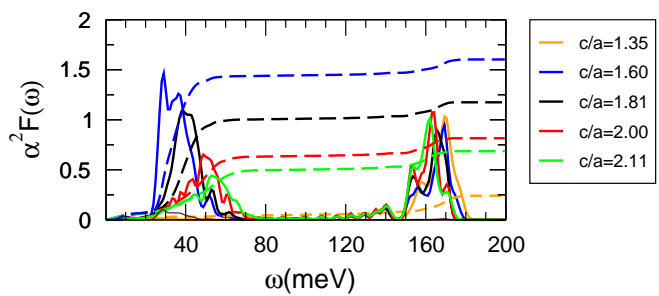}
\end{center}
\end{figure}

To gain a consistent picture of how the interplay of the three factors
affects the e-ph interaction in JIG, we study the Eliashberg function and the
frequency-dependent $\lambda (\omega )$ as a function of $c$, with $a$ and $n$
fixed at their values in CaC$_{6}$. In the $c$-range considered,
 both the position and the shape of the high-frequency peak,
which involves the $\pi ^{\ast }$-$\pi ^{\ast }$ coupled bond-stretching modes, is
nearly independent of $c$. For the low-frequency peak, involving the IL-$\pi
^{\ast }$ coupled buckling modes, the situation is entirely different. 
For the three higher values $c/a$=1.81, 2.00, 2.11 which match those of real GICs (CaC$_{6}$, SrC$_{6}$ and BaC$_{6}$)~\cite{latticeNOTE} for which first-principles calculations are available,\cite{GIC:calandra:band} the contribution to $%
\lambda $ from the low-frequency peak decreases with increasing  $c$:
as the graphene layers are pulled apart, the overlap between the IL and 
$\pi ^{\ast }$ wavefunctions decreases, the e-ph matrix-element of Eq.~\ref{g}  gets smaller, the buckling modes get harder.
Below $c$=1.35$a$, on the other hand, the low-frequency peak is just absent, because, at $n$=2/3, the IL band is empty, and gives no contribution to $\lambda$; so at first, as $c$ increases
beyond 1.35$a,$ the e-ph interaction must increase, because, with the
growing size of the IL sheet, the density of states entering Eq.~\ref{DOS}
grows. As a consequence $\lambda$, and thus $T_c$, will (for $n$=2/3) have a 
maximum for some $1.35\!<\!c/a\!<1.81$, which we calculated around 1.6 and
is also shown in Fig.~\ref{fig:fig8}.

Our calculated JIG trend  for $\lambda$ matches that of the real compounds (see
Fig.~7 of
Ref.~\onlinecite{GIC:calandra:band}). Details are given in Table~\ref{table1}%
, which also shows that although this trend is the right one, the total $%
\lambda $ in JIG is consistently larger than that of the
corresponding GIC. For a discussion of the possible reasons
for this discrepancy, see the end of the previous section.

\begin{table}[tbh]
\begin{ruledtabular}
\begin{tabular}{lcccc}
                    &$c/a$  & E$_{IL}$-E$_F$& $N(0)$    & $\lambda$  \\
JIG                & 1.35  &  -0.48        &  0.04     &  0.2       \\
JIG                & 1.60  &   0.93        &  0.10     &  1.6      \\
JIG                & 1.81  &   1.44        &  0.12     &  1.1       \\
JIG                & 2.00  &   1.66        &  0.14     &  0.8       \\
JIG                & 2.11  &   1.81        &  0.16     &  0.7       \\
CaC$_6$ $^{\dag}$   & 1.81 &   1.25         & 0.13      &  0.83      \\
SrC$_6$ $\;^{\dag}$ & 2.00 &   1.66        & 0.14      &  0.54      \\
BaC$_6$ $^{\dag}$ &  2.11&   1.46        & 0.12      &  0.38      \\
\end{tabular}
\end{ruledtabular}
\caption{ Comparison of JIG with $n\!$=\negthinspace 2/3 (this work)
with real alkali-metal intercalated graphites with
$n\!$=\negthinspace 2/3 (Ref.~ \onlinecite{GIC:calandra:band}).
$E_{IL}$ is the energy of the bottom of the IL band (eV). $N\left(
0\right) $ is the total DOS in states/(eV $\!\!\cdot \!\!$ C spin).
The corresponding $\protect\alpha ^{2}F(\protect\omega )$ are in
Fig.~\protect\ref{fig:fig8}.\newline The data marked with $^{\dag }$ are
taken from Ref.~ \onlinecite{GIC:calandra:band}; their e-ph coupling
$\protect\lambda $ also includes intercalant-related modes, which
contribute almost half the total coupling. } \label{table1}
\end{table}

\section{Conclusion}

In this paper we have studied the e-ph properties of electron-doped
graphite, using a jellium model in which the excess electrons are
neutralized by a uniform positive background of charge.
We have shown that, despite its
simplicity, the proposed model identifies and explains an important source
of the e-ph interaction in real GICs, namely, the existence of an interlayer
(IL) three-dimensional electronic band, whose filling determines the
occurrence of a strong, otherwise dormant e-ph interaction between IL and $%
\pi ^{\ast }$ electrons, for which we provided a simple understanding in
terms of Wannier-like functions.
Our picture corrects the rough idea according to which, since IL states
are free-electron like, they must experience little or no interaction with
the electronic states and phonon modes related to the carbon layers.
We have shown that, on the contrary,
IL states can experience a strong interaction with the surrounding C lattice, 
and thus give sizable contribution to the e-ph coupling, as observed
in GICs,~\cite{CaAlSi:gianto:band} disilicides and metal-sandwich 
structures.~\cite{LiB:curtarolo:band,LiB:mazin:band}
Our results amount to a natural explanation for
the empirical correlation between the filling of the IL band and the
occurrence of superconductivity in the GICs, and nicely fit 
the observed correlation between T$_c$ and layer spacing in actual GICs.

Moreover,
since $\pi ^{\ast }$ and IL electronic bands as well as carbon out-of-plane
phonon modes are found in several other materials, from nanotubes~\cite%
{NFES_in_CNTs} to layered compounds,~\cite{CaAlSi:gianto:band} including the
recently proposed metal sandwich structures,~\cite{LiB:curtarolo:band,LiB:mazin:band} our
findings may represent an inspiration for further research on the
electron-phonon mechanism identified here.

\section{Acknowledgements}

\label{ACKN} We thank F. Mauri and S. Massidda for useful discussion, and
Igor I. Mazin for suggestions and a critical reading of our manuscript. One
of us (GBB) gratefully acknowledges partial financial support from MUR (the
Italian Ministry for University and Research) through COFIN2005.

\section*{Appendix A: Tight-binding Model}

\subsection*{Undistorted JIG}

\begin{figure}[t]
\begin{center}
\includegraphics*[width=8.6cm]{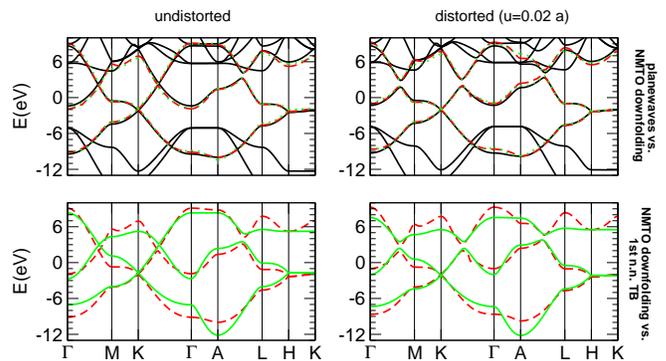}
\end{center}
\caption{(\emph{Color Online}) Band structures of undistorted (left)
and phonon-distorted (right) JIG (see text). Here the energy zero is
at the Fermi energy. Upper panels: comparison of the full band
structure based on
plane waves and pseudopotentials (black solid) with the three ($\protect\pi $%
, $\protect\pi ^{\ast }$ and IL) minimal downfolded bands (red dashed)
obtained from a (3x3) NMTO. As green dotted lines we also show the
corresponding $6^{th}$-n.n. tight-binding bands. Lower panels: we reproduce
the three relevant NMTO bands of the upper panels, and compare them with 1$%
^{st}$-n.n. first-principle tight-binding bands (green solid). These bands
look obviously poorer than those with 6 neighbors, but they still correctly
reproduce the band splittings due to the phonon displacement (see text).
}
\label{fig:fig9}
\end{figure}

The NMTO method~\cite{Theory:andersen:TBLMTOASA,
Theory:andersen:NMTO} allows us to define from first principles
three most localized ``atomic'' orbitals which span the Hilbert
space of the electronic states corresponding to three bands which,
in the main paper, were labeled as $\pi$, $\pi^*$, IL, and
associated to the e-ph interaction in GICs. What do they look like
and where do they sit? To answer this question we need the geometry
of our JIG model. ``Primitive graphite'' C$_2$ has
$\alpha\alpha\alpha$ stacking; direct and reciprocal lattices are
defined by the basis vectors
\begin{equation*}
\left[
\begin{array}{c}
\mathbf{T}_{1} \\
\mathbf{T}_{2} \\
\mathbf{T}_{3}%
\end{array}%
\right] =\left[
\begin{array}{ccc}
\frac{a\sqrt{3}}{2} & \frac{a}{2} & 0 \\
0 & a & 0 \\
0 & 0 & {c}%
\end{array}%
\right] ; \left[
\begin{array}{c}
\mathbf{G}_{1} \\
\mathbf{G}_{2} \\
\mathbf{G}_{3}%
\end{array}%
\right] =2\pi \left[
\begin{array}{ccc}
\frac{2}{a\sqrt{3}} & 0 & 0 \\
-\frac{1}{a\sqrt{3}} & \frac{1}{a} & 0 \\
0 & 0 & \frac{1}{c}%
\end{array}
\right] \, ;
\end{equation*}

the positions of the two C atoms within the unit cell are%
\begin{equation*}
\pm \mathbf{C\equiv }\pm \frac{1}{3}\left( \frac{\mathbf{T}_{1}}{2}-\frac{%
\mathbf{T}_{2}\!-\!\mathbf{T}_{1}}{2}\right) =\left( \frac{\pm a}{2\sqrt{3}}%
,0,0\right) \,,
\end{equation*}%
and the position of the interstitial site \textbf{S}, where the intercalant
may sit in real GICs, is:
\begin{equation*}
\mathbf{S\equiv }\frac{1}{2}\mathbf{T}_{3}=\left( 0,0,\frac{c}{2}\right) .
\end{equation*}%
Out of our three localized orbitals, two sit on the two carbon atoms of the C%
$_{2}$ unit cell, are of $p_{z}$ symmetry, and are naturally identified with
the carbon $p_{z}$ orbitals; the third sits on the interstitial site and,
although it looks like an $s$ orbital (see Fig.~\ref{fig:fig2}), has only
axial symmetry (around $z$); for simplicity of notation, however, we assign
it an $s$ subscript, keeping in mind that this is \emph{not} a purely $s$
orbital:
\begin{equation*}
\left\langle \mathbf{r}|\pm \mathbf{C}\right\rangle =\psi _{p_{z}}(\mathbf{r}%
\mp \mathbf{C})\quad ;\quad \left\langle \mathbf{r}|\mathbf{S}\right\rangle
=\psi _{s}(\mathbf{r}-\mathbf{S})\,.
\end{equation*}%
If we diagonalize the electronic hamiltonian $\hat{h}$ in the $3\!\times \!3$
basis of the localized states $|+\mathbf{C}\rangle $, $|-\mathbf{C}\rangle $%
, and $|\mathbf{S}\rangle $, or, more precisely, of the three corresponding
Bloch sums
\begin{eqnarray}
\left\vert {\pm {\mathbf{C,k}}}\right\rangle &=&\sum_{\mathbf{T}}\mathrm{e}%
^{i\mathbf{k\cdot T}}\left\vert \mathbf{T}\!\pm \!\mathbf{C}\right\rangle \\
\left\vert {\mathbf{S},\mathbf{k}}\right\rangle &=&\sum_{\mathbf{T}}\mathrm{e%
}^{i\mathbf{k\cdot T}}\left\vert \mathbf{{T}+{S}}\right\rangle  \label{a4}
\end{eqnarray}%
where the sum extends to all lattice vectors $\mathbf{T}$, we
recover, by construction, the three electronic bands ($\pi $, $\pi
^{\ast }$ and IL) from which these localized states were derived,
and obtain the corresponding Bloch states ($|\pi ;\mathbf{k}\rangle
,|\pi ^{\ast };\mathbf{k}\rangle $ and
$|\mathrm{IL};\mathbf{k}\rangle $) as a linear combination of these
three Bloch sums. The three relevant NMTO bands are shown as red
dashed lines in the upper panels of Fig.~\ref{fig:fig9} for our JIG
without (left) and with (right) a $\Gamma $-point buckling phonon
displacement of 0.17 a.u.; the full JIG band structure from
pseudopotentials and plane waves is shown as black solid lines, for
comparison. With our $3\!\times \!3$ basis, an essentialy perfect
\textquotedblleft first-principles\textquotedblright\ tight-binding
representation is obtained by truncating the electronic hamiltonian
matrix beyond the sixth shell of nearest neighbors (n.n.), and shown
by the green dotted lines in the upper panels of Fig~\ref{fig:fig9}. The
corresponding formulae and parameters are available upon request.~\cite%
{askLilia}

\begin{table}[b!]
\caption{$3\!\times\!3$ nearest-neighbor first-principles TB
parameters for undistorted JIG, from NMTO (see text and
Table~\protect\ref{TBundist}), in eV. Unlike
Fig.~\protect\ref{fig:fig9}, here the energy zero is at the crossing
point of the $\protect\pi$ and $\protect\pi^*$ bands, so E$_F=$ 2.0
eV. A carbon-buckling phonon of amplitude $u=0.02 \, a$ (where $a$
is the planar lattice constant of C$_2$) changes $t_{sp}$ by as much
as $0.1122$ eV. To
linear order this means $\protect\delta t_{sp}/\protect\delta u=2.241$ eV/%
\AA .\newline
}
\label{tableTB}%
\begin{ruledtabular}
\begin{tabular}{rr}
$\varepsilon_{s}-\varepsilon_{p}\phantom{-}= \phantom{-}3.5115$     &
$t_{sp}= \,\,0.8155$      \\
$\theta_s=\theta_s^{xy}\,\,=-0.8811$
   & $t_{pp}\,\,=\,\,t_{pp}^{xy}= \,\,2.5626$
\end{tabular}
\end{ruledtabular}
\end{table}

\begin{table*}[t!]
\caption{The $3\!\times\!3$ electronic hamiltonian for JIG when only
nearest neighbor integrals (see Table~\protect\ref{tableTB} and
text) are retained. }
\label{TBundist}%
\begin{tabular}{ccccccc}
&  & $\left| \mathbf{S};\mathbf{k}\right\rangle$ &  & $\left| \mathbf{C};%
\mathbf{k} \right\rangle$ &  & $\left| - \mathbf{C};\mathbf{k}\right\rangle$
\\
&  &  &  &  &  &  \\
$\left\langle \mathbf{S};\mathbf{k}\right|$ &  & $(\varepsilon_{s}\!-\!%
\varepsilon_{p}) +2\theta_{s} \left[ \cos(2\lambda)+\cos(2\mu)+\cos(2\nu)%
\right]$ &  & $-2i e^{-i\xi}t_{sp}\sin(\xi) \left(
1+e^{-2i\mu}\!\!+e^{-2i\nu}\right)$ &  & $-2i e^{-i\xi}t_{sp}\sin(\xi)
\left( 1+e^{-2i\mu}\!\!+e^{-2i\lambda}\right)$ \\
&  &  &  &  &  &  \\
$\left\langle \mathbf{C};\mathbf{k}\right|$ &  & $2i e^{i\xi}t_{sp}\sin(\xi)
\left( 1+e^{2i\mu}+e^{2i\nu}\right)$ &  & 0 &  & $-t_{pp}(1+e^{2i%
\lambda}+e^{-2i\nu})$ \\
&  &  &  &  &  &  \\
$\left\langle \mathbf{- C};\mathbf{k}\right|$ &  & $2i
e^{i\xi}t_{sp}\sin(\xi) \left( 1+e^{2i\mu}+e^{2i\lambda}\right)$ &  & $%
-t_{pp}(1+e^{-2i\lambda}+e^{2i\nu})$ &  & 0 \\
&  &  &  &  &  &  \\
&  &  &  &  &  &
\end{tabular}%
\end{table*}

However, to understand the basic features of the band structure and e-ph
interaction, nearest-neighbor matrix elements may be enough. In this case
only a few integrals survive: if $\alpha$ and $\beta$ equal $s$ for the $|
\mathbf{S} \rangle$ orbital which sit on the interstitial site and $p$ for
the two carbon $| \pm \mathbf{C} \rangle$ orbitals, we have onsite hopping
integrals indicated by $\varepsilon_\alpha$ and $\varepsilon_\beta$, hopping
integrals between orbitals of types $\alpha,\beta$ sitting on n.n. \textit{%
atoms} indicated by $t_{\alpha\beta}$ (and all taken \emph{positive}, $%
t\equiv \left| t\right|$), and hopping integrals between orbitals of type $%
\alpha$ sitting on the same atom of n.n. \textit{unit cells} indicated by $%
\theta_\alpha$. Their numerical values are in Table~\ref{tableTB}.

\begin{figure}[b]
\begin{center}
\includegraphics*[width=4cm]{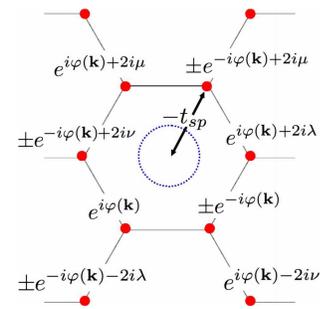}
\end{center}
\caption{(\emph{Color Online}) Schematic real-space picture of the three
Bloch sums $|\mathbf{S}; \mathbf{k} \rangle$ and $|\protect\pi_{\pm};
\mathbf{k} \rangle$ in a single graphitic plane (see text). Red dots mark
the position of carbon atoms; the blue dotted circle represents the
interstitial localized orbital $|\mathbf{S} \rangle$ sitting in the central
unit cell; similar wavefunctions are at the center of all other unit cells,
multiplied by an appropriate phase factor, but are not shown here. In the
vertical (out-of-plane) direction, wavefunctions are multiplied by an
additional phase factor $e^{2i\protect\xi}$. }
\label{fig:fig10}
\end{figure}

Then the hamiltonian matrix is given by Table~\ref{TBundist}, where
\begin{eqnarray}
&\lambda\equiv \mathbf{k}\cdot {\displaystyle \frac{\mathbf{T}_{1}}{2}}={%
\displaystyle \frac{a}{4}}\left( k_{y}\!+\!\sqrt{3} k_{x}\right) ,\quad
\mu\equiv \mathbf{k}\cdot {\displaystyle \frac{\mathbf{T}_{2}}{2}}={%
\displaystyle \frac{a}{2}}\, k_{y} \, , &  \notag \\
&\nu\equiv \mathbf{k}\cdot {\displaystyle \frac{\mathbf{T}_{2}\!-\!\mathbf{T}%
_{1}}{2}}=\mu-\lambda={\displaystyle \frac{ a}{4}}\left( k_{y}\!-\!\sqrt{3}%
k_{x}\right) , &  \notag \\
&\xi\equiv \mathbf{k}\cdot {\displaystyle \frac{\mathbf{T}_{3}}{2}}={%
\displaystyle \frac{c}{2}}\,k_{z} \,; &  \notag
\end{eqnarray}
the corresponding n.n. bands are shown as solid green lines in the lower
panels of Fig.~\ref{fig:fig9}, where they are compared to the original NMTO
bands (red dashed). Notice that, in this model, the $k_z$ dispersion
entirely comes from the $s\!-\!p$ hopping between the $s$ orbital sitting on
the interstitial site and the carbon $p_z$ orbitals on the graphite planes
immediately above and below it: carbon $p\!-\!p$ matrix elements between
adjacent graphite planes are beyond n.n. In the $k_z\!\!=\!\!0$ plane ($%
\xi\!\!=\!\!0$) the matrix element between $s$ and $p_z$ orbitals is
identically zero, because the phases of the $p_z$ orbitals on two adjacent
graphene sheets cancel. Therefore, for $k_z\!\!=\!\!0$, the IL band is
\begin{eqnarray}
\varepsilon_{\mathrm{IL}}(\mathbf{k}) = \varepsilon_{s}+2\theta_{s} \biggl[ %
\cos\left(\frac{k_ya+\sqrt{3}k_xa}{2}\right)+ \cos k_ya  \notag \\
+ \cos\left(\frac{k_ya-\sqrt{3}k_xa}{2}\right) \biggr]\, ,
\end{eqnarray}
and its eigenvector $|\mathrm{IL};\mathbf{k}\rangle=|\mathbf{S};\mathbf{k}%
\rangle$ just coincides with the Bloch sum of Eq.~\ref{a4}; the $\pi,\pi^*$
bands entirely derive from the $p_z$ sub-block, whose diagonalization
yields:
\begin{eqnarray}
\varepsilon _{\pi^*, \pi} (\mathbf{k}) = \varepsilon _{\pi \mp } (\mathbf{k}%
) = \varepsilon _{p} \pm t_{pp}\left\vert
1+e^{2i\lambda}+e^{-2i\nu}\right\vert =  \label{a6} \\
=\varepsilon _{p} \pm t_{pp}\sqrt{3+4\cos \frac{k_{y}a}{2}\cos \frac{\sqrt{3}%
k_{x}a}{2}+2\cos k_{y}a}  \notag
\end{eqnarray}%
with eigenvectors
\begin{eqnarray}
\left\vert \pi^{*}; \mathbf{k}\right\rangle&=&\left\vert \pi_{-}; \mathbf{k}%
\right\rangle =\frac{1}{\sqrt{2}}\left\{ \left\vert -\mathbf{C};\mathbf{k}%
\right\rangle e^{i\varphi \left( \mathbf{k}\right) }-\left\vert \mathbf{C};%
\mathbf{k}\right\rangle e^{-i\varphi \left( \mathbf{k}\right) }\right\}
\notag \\
\left\vert \pi\phantom{^{*}}; \mathbf{k}\right\rangle &=& \left\vert
\pi_{+}; \mathbf{k}\right\rangle =\frac{1}{\sqrt{2}}\left\{ \left\vert -%
\mathbf{C};\mathbf{k}\right\rangle e^{i\varphi \left( \mathbf{k}\right)
}+\left\vert \mathbf{C};\mathbf{k}\right\rangle e^{-i\varphi \left( \mathbf{k%
}\right) }\right\} ,\quad  \notag
\end{eqnarray}
where $2\varphi \left( \mathbf{k}\right) $ is the phase of $\left(
1+e^{2i\lambda}+e^{-2i\nu}\right) $. The state $\left\vert \pi
_{+}\right\rangle$ has lower eigenvalue and is of bonding character, being a
bonding combination of the carbon ${p_z}$ orbitals at $\mathbf{k}=0$; the
other one, $\left\vert \pi _{-}\right\rangle$, is the antibonding
combination. In Fig.~\ref{fig:fig10} we show a real-space cartoon of these three
complex Bloch sums, only in the $xy$ plane; in the vertical direction (not
shown) both Bloch sums $|\pi_\pm ;\mathbf{k}\rangle$ are odd with respect to
the plane, since both derive from $p_z$ states; orbitals on different planes
differ by a phase factor $e^{2i\xi}$.
\begin{figure}[tbp]
\begin{center}
\includegraphics*[width=4cm]{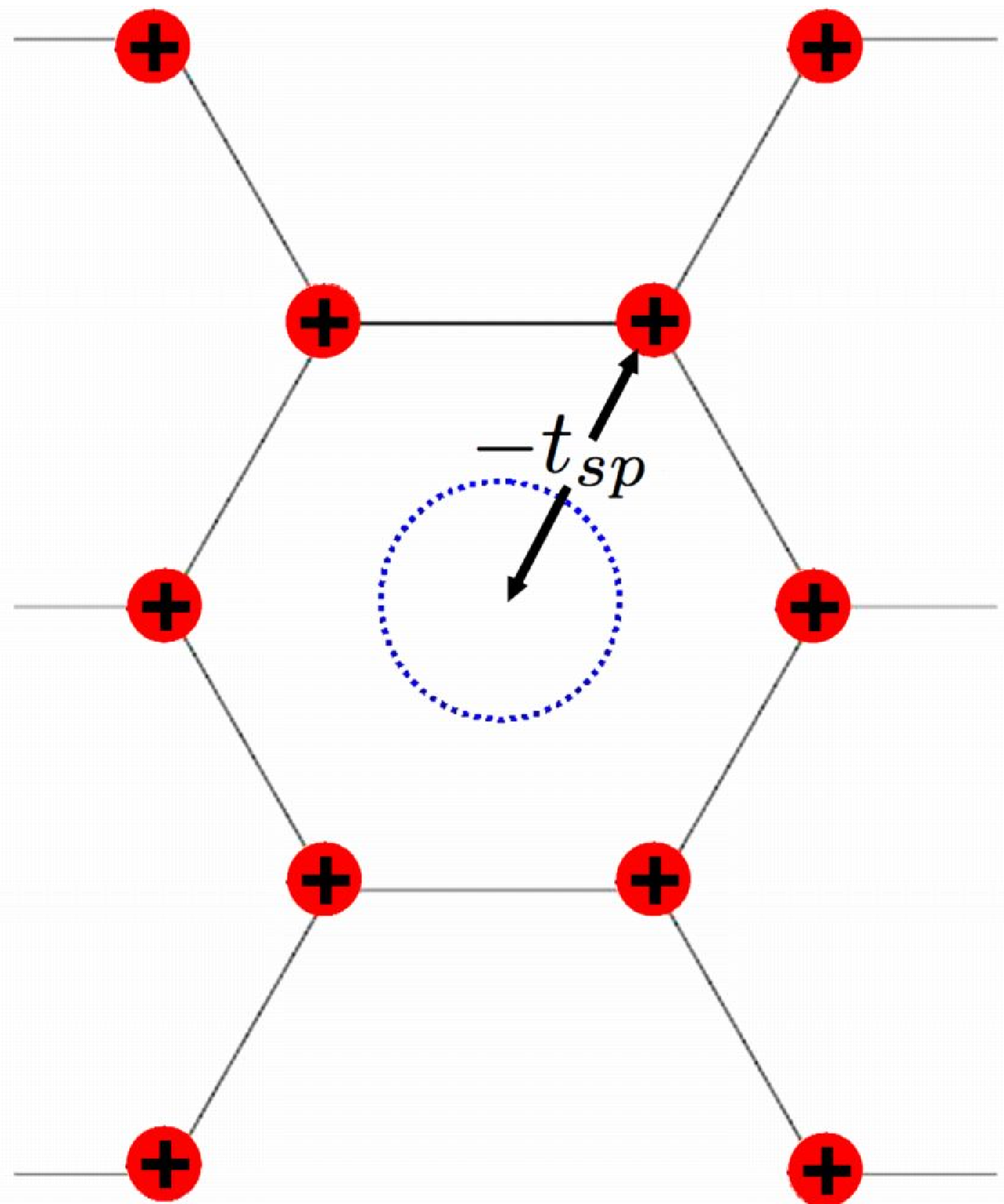} 
\includegraphics*[width=4cm]{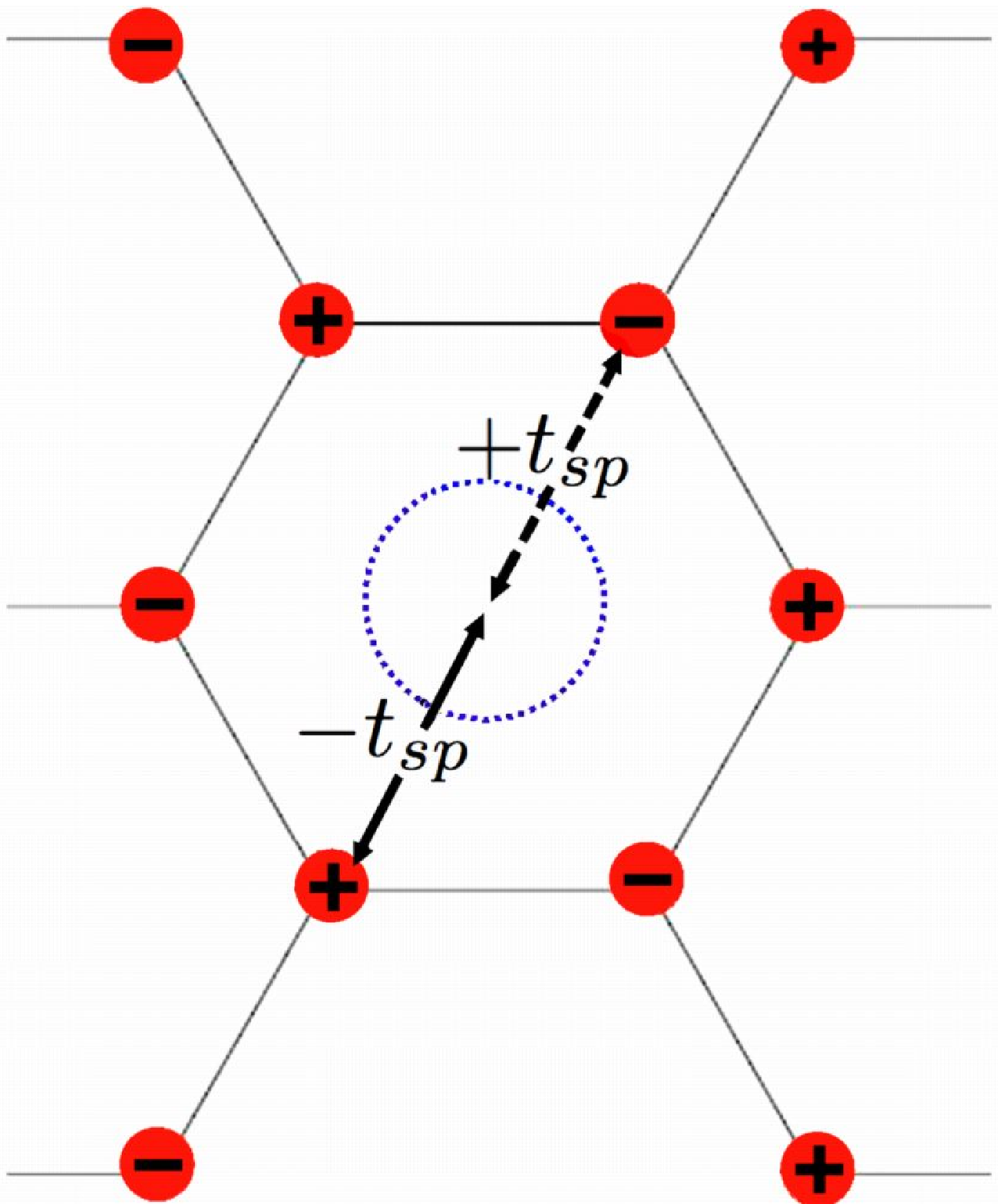} 
\end{center}
\caption{(\emph{Color Online}) Schematic real-space picture, in the $xy$
plane, of the Bloch sums $|\mathbf{S}; \mathbf{k} \rangle$, $|\protect\pi%
_{+}; \mathbf{k} \rangle$ (left), and $|\protect\pi_{-}; \mathbf{k} \rangle$
(right), for $k_x\!=\!k_y\!=\!0 \rightarrow \protect\varphi(\mathbf{k})=0$ ($%
\Gamma-A$ line). }
\label{fig:fig11}
\end{figure}

For $k_z\!\!\ne 0\!\!$ \, the matrix element between $s$ and $p_z$ states is
not zero, so, unlike the $k_z\!\!= 0\!\!$ \, case, the $|\mathbf{S};\mathbf{k%
}\rangle$ and $|\pi_\pm ;\mathbf{k}\rangle$ Bloch sums may hybridize. A
priori this means that the matrix elements of the electronic hamiltonian $%
\langle \pi_{\pm}; \mathbf{k} | \hat h | \mathbf{S}; \mathbf{k} \rangle$ may
be nonzero; however it turns out that one of the two, the antibonding Bloch
sum, gives $\langle \pi_{-}; \mathbf{k} | \hat h | \mathbf{S}; \mathbf{k}
\rangle = 0$ almost everywhere. The only exception is a tiny $\mathbf{k}$%
-space region between the $A$ and $L$ points; the corresponding, small gap
between the $\pi^{*}$ and the IL bands is visible in Fig.~\ref{fig:fig9} but not
in Fig.~\ref{fig:fig1}, because of the thickness of the ``fat bands''. The
reason why $\langle \pi_{-}; \mathbf{k} | \hat h | \mathbf{S}; \mathbf{k}
\rangle = 0$ may be immediately understood along the $\Gamma\!-\!A$ line ($%
k_x\!=\!k_y\!=\!0$), where $\varphi(\mathbf{k})\!=\!0$ (see Fig.~\ref{fig:fig11},
where we display the three corresponding Bloch sums): each $\mathbf{S}$ site
has 12 carbon nearest neighbors, 6 above and 6 below, which respectively
carry, in the vertical direction, phase factors of 1 and $e^{2i\xi}$, so for
the bonding ($+$) and antibonding ($-$) Bloch sums we obtain:
\begin{eqnarray}
\langle \pi_{+}; \mathbf{k} | \hat h | \mathbf{S}; \mathbf{k} \rangle &=& -%
\frac{1}{\sqrt{2}}\left(6 t_{sp} + 6t_{sp}e^{2i\xi} \right)  \notag \\
\langle \pi_{-}; \mathbf{k} | \hat h | \mathbf{S}; \mathbf{k} \rangle &=& -%
\frac{1}{\sqrt{2}}\!\left[ 3 t_{sp} - 3 t_{sp} + ( 3 t_{sp} - 3
t_{sp})e^{2i\xi} \right]=0  \notag
\end{eqnarray}
However, even for any other $\mathbf{k}$ point in the Brillouin zone (BZ) a
similar, just more lengthy procedure, based on the phases of Fig.~\ref{fig:fig10}%
, gives zero for the antibonding Bloch sum; except, as mentioned, in a small
region between the $A$ and $L$ points, where the definition of bonding and
antibonding looses its meaning. In other words, the electronic states
corresponding to the $\pi^*$ band may legitimately be identified with the $%
|\pi_{-} ;\mathbf{k}\rangle$ Bloch sum of $p_z$ orbitals over almost the
entire BZ, with no or very little admixture of the Bloch sum of the
interstitial orbitals $|\mathbf{S}; \mathbf{k} \rangle$.

\subsection*{Distorted JIG}

We have seen that, in the undistorted lattice, the matrix element of the
electronic hamiltonian between the Bloch sum of interstitial orbitals $|%
\mathbf{S}; \mathbf{k} \rangle$ and the antibonding Bloch sum $|\pi_{-};
\mathbf{k} \rangle$ is almost always zero, so that the latter can be
practically identified as the Bloch state of the $\pi^*$, while the Bloch
state of the IL band will have a prevailing $|\mathbf{S}; \mathbf{k} \rangle$
character plus, at certain $\mathbf{k}$'s, some admixture of the bonding $%
|\pi_{+}; \mathbf{k} \rangle$ Bloch sum. Upon out-of-plane distortions of
the graphene sheets, our e-ph \textit{ab-initio} calculations reveal,
instead (see main paper), a large coupling between the the $\pi^*$ band and
the IL band. How do we understand that? By symmetry: in graphene the
relevant out-of-plane phonons have the same symmetry as the $\pi^*$
electronic states. The relevance of this symmetry argument appears if we
estimate the e-ph coupling between the electronic states $i$ and $j$ due to
the phonon mode ($\nu \mathbf{q}$) with the change of the corresponding
electronic matrix element upon phonon displacement
\begin{eqnarray}
\left\langle i, \mathbf{k} \right\vert \delta \hat h_{\nu \mathbf{q}}
\left\vert j, \mathbf{k+q}\right\rangle ,  \notag
\end{eqnarray}
multiplied by the nesting function $\delta (\varepsilon_{i\mathbf{k}})\delta
(\varepsilon_{j\mathbf{k+q}})$ and integrated over the Fermi surface. Within
our nearest-neighbor model this well defined, but not immediate procedure,
ultimately recovers the high coupling obtained in the full first-principles
calculation (see main paper). But at special points in the BZ our model
allows to understand the underlying symmetry argument by inspection. The
simplest example is the $\Gamma$ optical buckling phonon ($\mathbf{q}\!=\!0$%
). This is the frozen-in distortion considered in the right panels of Fig.~%
\ref{fig:fig9}, whose most visible effect on the electronic states is the
appearance of a hybridization gap between the $\pi^*$ and IL bands at $\sim
2 $ eV above the Fermi energy, in the middle of the $\Gamma\!-\!K$ and $%
\Gamma\!-\!M$ lines. If, upon a $\mathbf{q}\!=\!0$ phonon distortion, the
electronic hamiltonian is written as $\hat{h}_u=\hat{h}_{u=0}+\delta \hat{h}%
_{u}$, then for a general $\mathbf{k}\!\ne\!0$ the change in the electronic
matrix element relevant to the e-ph coupling becomes $\left\langle i,
\mathbf{k} \right\vert \delta \hat{h}_{u} \left\vert j, \mathbf{k}%
\right\rangle$ and is given in Table~\ref{TBdist}. From it one may already
get some intuition of the strong coupling between the $\pi^*$ band and the
IL band.
\begin{table*}[t]
\begin{tabular}{ccccccc}
&  & $\left| \mathbf{S};\mathbf{k}\right\rangle $ &  & $\left| \mathbf{C};%
\mathbf{k}\right\rangle $ &  & $\left| - \mathbf{C};\mathbf{k}\right\rangle$
\\
&  &  &  &  &  &  \\
$\left\langle \mathbf{S};\mathbf{k}\right| $ &  & 0 &  & $2 e^{-i\xi} \delta
t_{sp}\cos(\xi)\left( 1+e^{-2i\mu}+e^{-2i\nu}\right) $ &  & $- 2
e^{-i\xi}\delta t_{sp}\cos(\xi)\left( 1+e^{-2i\mu}+e^{-2i\lambda}\right)$ \\
&  &  &  &  &  &  \\
$\left\langle \mathbf{C};\mathbf{k}\right|$ &  & $2 e^{i\xi} \delta t_{sp}
\cos(\xi)\left( 1 + 1+e^{2i\mu}+e^{2i\nu}\right)$ &  & 0 &  & 0 \\
&  &  &  &  &  &  \\
$\left\langle \mathbf{- C};\mathbf{k}\right| $ &  & $- 2 e^{i\xi}\delta
t_{sp}\cos(\xi)\left( 1+e^{2i\mu}+e^{2i\lambda}\right) $ &  & 0 &  & 0 \\
&  &  &  &  &  &  \\
&  &  &  &  &  &
\end{tabular}%
\caption{Change in the JIG nearest-neighbor tight-binding electronic
hamiltonian matrix elements under a $\Gamma$-buckling-phonon
distortion (see text and Tables~\protect\ref{tableTB} and
\protect\ref{TBundist}). } \label{TBdist}
\end{table*}

\begin{figure}[tbp]
\begin{center}
\includegraphics*[width=4cm]{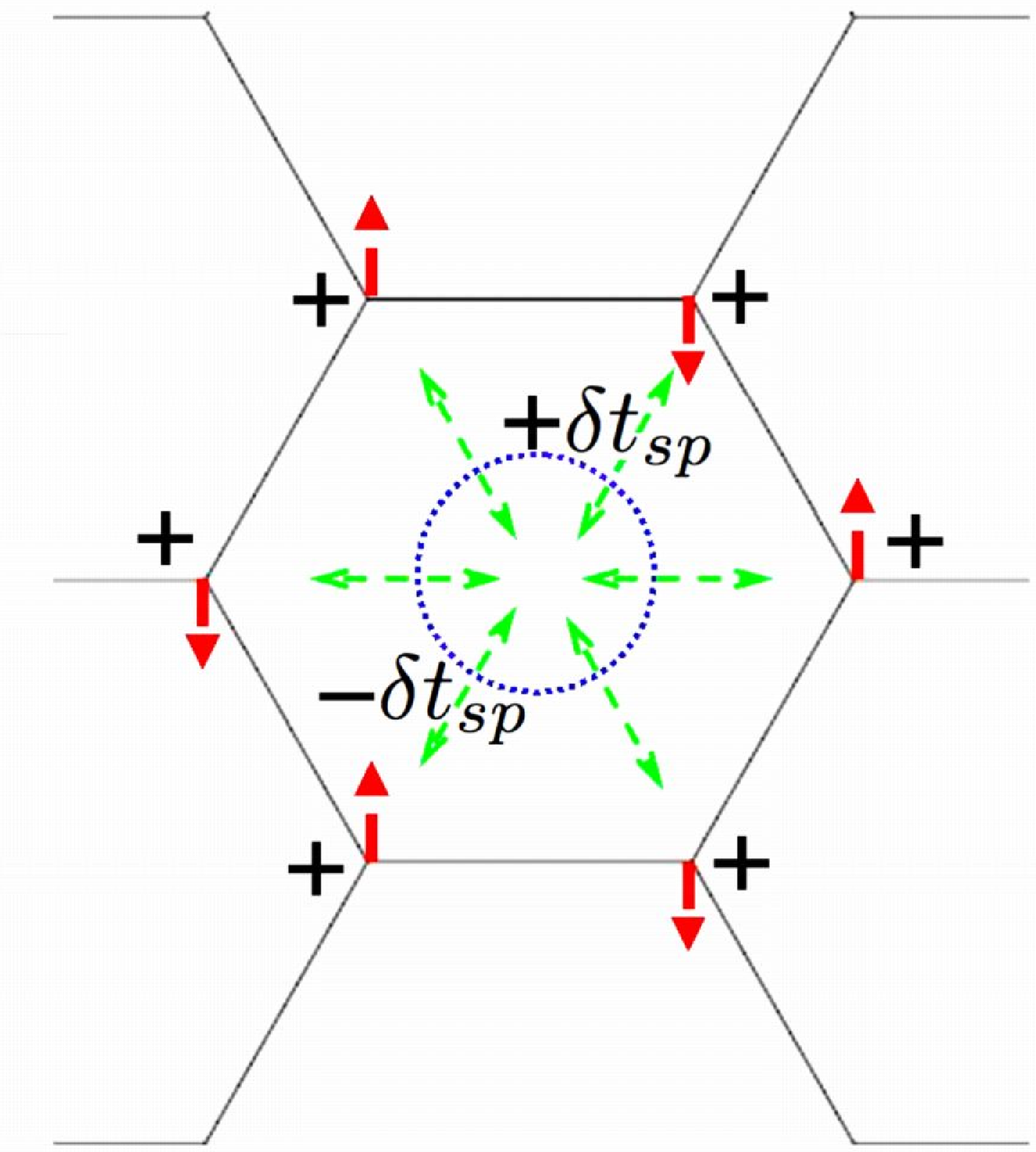}
\includegraphics*[width=4cm]{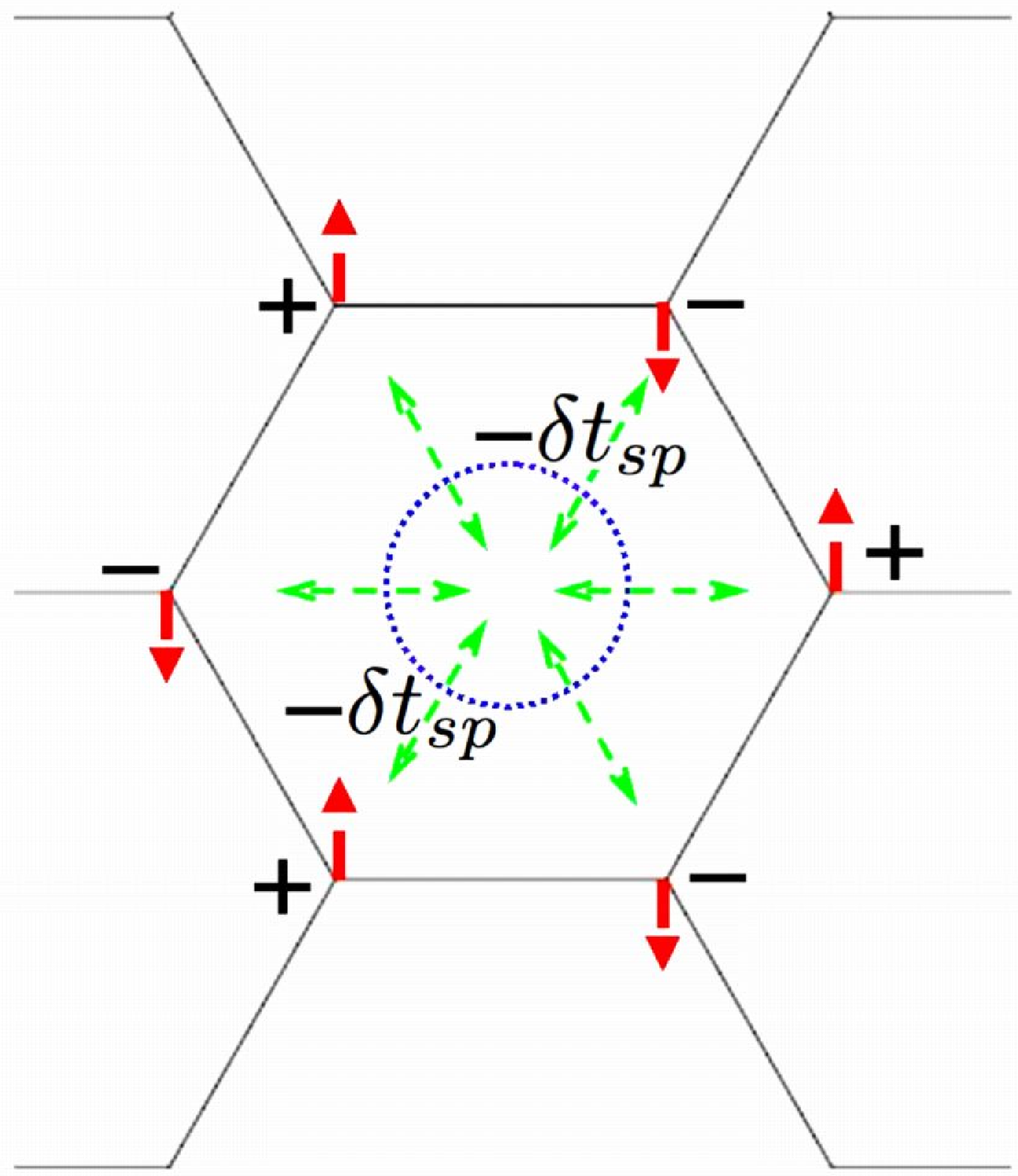}
\end{center}
\caption{(\emph{Color Online}) Schematic picture, in the $xy$ plane, of the
Bloch sums $|\mathbf{S}; \mathbf{k} \rangle$, $|\protect\pi_{+}; \mathbf{k}
\rangle$ (left), and $|\protect\pi_{-}; \mathbf{k} \rangle$ (right), at the $%
\Gamma$ point of graphene with an optical buckling distortion of $\mathbf{q}%
\!=\!0$. }
\label{fig:fig12}
\end{figure}

At $\mathbf{k}=0$ the symmetry argument becomes obvious: the $\delta \hat
h_{\nu \mathbf{q}} $ induced by the optical buckling phonon (alternate
raising and lowering of carbon atoms) has the same symmetry as the $|
\pi^{*}; \mathbf{k}\!=\!0\rangle=| \pi_{-}; \mathbf{k}\!=\!0\rangle$
electronic state (alternate sign on carbon atoms, antibonding combination of
$p_z$ orbitals); the product of two anti-bonding functions is bonding; as a
result, we have a strong overlap with the IL electronic state, which, at $%
\mathbf{k}\!=\!0$, is the simple sum of interstitial $|\mathbf{S}\rangle$
orbitals. This is schematically shown in Fig~\ref{fig:fig12}. The only
difference with respect to the purely electronic analysis of the previous
cartoons (Figs.~\ref{fig:fig10}, \ref{fig:fig11}) is that, besides the electronic
phases, we must now consider the phonon phases too. Here $\delta t_{sp}$
indicates the change of the corresponding matrix element upon distortion;
quantitatively, for a distortion amplitude $u=0.17$ a.u., the $t_{sp}$
matrix element changes by $\delta t_{sp}=0.101$ eV. For those carbon atoms
which move towards the interstitial site, where $|\mathbf{S}\rangle$ sits,
the matrix element becomes more negative: $-t_{sp} \to -(t_{sp} + \delta
t_{sp})$; for those which move away from it, it becomes less negative: $%
-t_{sp} \to -(t_{sp} - \delta t_{sp})$. We then have, respectively for the
bonding and antibonding Bloch sums:
\begin{eqnarray}
&&\langle \pi_{+}; \mathbf{k}\!=\!0 | \delta \hat h_u | \mathbf{S}; \mathbf{k%
}\!=\!0 \rangle =  \notag \\
&&- \frac{2 \delta t_{sp}}{\sqrt{2}} \left[ 1 + (-1) + 1 + (-1) + 1 + (-1) %
\right] = 0 \, ;  \notag \\
&&\langle \pi_{-}; \mathbf{k}\!=\!0 | \delta \hat h_u | \mathbf{S}; \mathbf{k%
}\!=\!0 \rangle =  \notag \\
&& - \frac{2 \delta t_{sp}}{\sqrt{2}} \left[ 1 - (-1) + 1 - (-1) + 1 - (-1) %
\right] = - \frac{12}{\sqrt{2}}\delta t_{sp} \, .  \notag
\end{eqnarray}
In conclusion, with our simple TB we obtain, at first glance for the $\Gamma$
point, and with some additional work at any point in the BZ, a simple
physical picture not only for the electronic states, as seen in Figs.~\ref%
{fig:fig10} and~\ref{fig:fig11}, but also for the e-ph interaction.

\section*{Appendix B: Computational Details}

The results presented here were obtained within
density functional theory~\cite{DFT} using two different \emph{ab-initio}
methods: for \textquotedblleft fat bands\textquotedblright , Wannier-like
orbitals, and tight-binding parameters, we used TB-LMTO-ASA~\cite%
{Theory:andersen:TBLMTOASA} and NMTO,~\cite{Theory:andersen:NMTO} and for
total energies, phonon frequencies and e-ph couplings, we employed
plane-waves and pseudopotentials.~\cite{PWscf} All methods give the same
results for the electronic bands, the Fermi surface, and several other
properties, as was carefully cross-checked.

For the self-consistent calculation of the band structure, Fermi surfaces,
phonon dispersions and e-ph coupling~\cite{PWscf} of jellium-doped graphite
we employed ultra-soft carbon pseudopotentials,~\cite{Vanderbilt} with a
cutoff or 30 (360) Ryd for the wave-functions (density) respectively, and
PBE exchange-correlation functional.~\cite{DFT:PBE} For the $\mathbf{k}$%
-space integration of the self-consistent calculations we employed a $16
\times 16 \times 8$ Monkhorst-Pack grid,~\cite{MPgrid} yielding 150 points
in the IBZ, with a 0.06 Ryd cold smearing;~\cite{Marzari:coldsmearing} a
much denser $32 \times 32 \times 16$ grid was used for the integration on
the Fermi surface of the e-ph matrix elements. This led to a convergence of $%
\mathrm{\sim}1$ meV for the phonon frequencies, except for the acoustical
out-of-plane branch along the $\Gamma\!-\!A$ line, which requires a much
higher cutoff to converge.
For consistency check, we also calculated the $\Gamma-$point frequencies
of graphite in the experimental structure, obtaining frequencies within
$\pm 0.1\,$meV of Ref~\onlinecite{graf:marzari:band}.
 Dynamical matrices and phonon linewidths were
calculated on a $16 \times 16 \times 8$ grid in $\mathbf{q}$-space and the
phonon dispersions were then obtained by Fourier interpolation. Such a dense
mesh in $\mathbf{q}$-space was necessary to resolve the effect of the Fermi
surface nesting on the e-ph coupling (Fig.~\ref{fig:fig3}).

For the fat bands and Wannier-like orbitals, we used the TB-LMTO and NMTO
methods.~\cite{Theory:andersen:TBLMTOASA,Theory:andersen:NMTO} In this case,
to obtain a good representation of the electronic wavefunction both in the
neighborhood of C atoms and in the interstitial regions, we had, in our C$_2$
unit cell, an $s,p,d$ basis set on each C atom, but also two ``empty
spheres'' (with $s,p$ channels): one, small, in the C plane; another, very
large, between the planes, on the interstitial site called $\mathbf{S}$ in
appendix A, whose crystal coordinates are $(0,0,\frac{c}{2})$. This is where
the intercalant may sit in real GICs: for example, on each of these
interstitial sites in LiC$_2$; on every third such sites in CaC$_6$.


\begin{thebibliography}{99}
\bibitem{NMTOnote} With the NMTO downfolding method we obtain a minimal
basis set which spans the three $\pi $, $\pi ^{\ast }$ and IL bands as shown
red in the upper part of Fig. 8. This basis consists of two C$\,p_{z}$ and
one E$\,s$ $N$th-order muffin-tin orbital (NMTOs) per primitive cell. What
is shown in Fig.~\ref{fig:fig10} is the latter orbital. It is localized, $s$%
-like, and centered on an interstital site (E). After symmetrical
orthonormalization to all other basis funtions in the minimal set, it
becomes the orbital named $|\mathbf{S}\rangle $ in Appendix A.
\bibitem{lambdaNOTE} In our total $\lambda $, we exclude the (negligible)
contributions arising from the imaginary frequencies along
 $\Gamma \!-\!A\!-\!L$.
\bibitem{latticeNOTE} In real CaC$_{6}$, SrC$_{6}$ and BaC$_{6},$ $a$
changes by less than 1$\%,$ while $c$ changes by 20\%. This tiny change in $%
a $ changes the energy of the buckling phonons by less than 0.1 meV, which
is negligible, and hardens the in-plane phonons by $\mathrm{\sim}5\,$meV.
\end{thebibliography}
\end{document}